\documentclass[aps,prd,onecolumn,groupedaddress,showpacs,nofootinbib,amssymb]{revtex4}
\usepackage[dvips]{graphicx}
\usepackage{amssymb}
\usepackage{amsmath}
\usepackage{graphicx,,color}
\usepackage{amsfonts}
\usepackage{bm}
\usepackage{cancel}
\usepackage{comment}
\usepackage[dvipsnames]{xcolor}
\usepackage{makecell}

\allowdisplaybreaks[4]

\begin{document}

\tolerance=5000

\title{A Panorama of Viable $F(R)$ Gravity Dark Energy Models}
\author{V.K.
Oikonomou,$^{1,2}$}\email{v.k.oikonomou1979@gmail.com;voikonomou@auth.gr}\author{Ifigeneia
Giannakoudi$^1$}\email{ifigeneiagiannakoudi@gmail.com}
\affiliation{$^{1)}$Department of Physics, Aristotle University of Thessaloniki, Thessaloniki 54124, Greece\\
$^{2)}$ Laboratory for Theoretical Cosmology, International Center
of Gravity and Cosmos, Tomsk State University of Control Systems
and Radioelectronics  (TUSUR), 634050 Tomsk, Russia}

\tolerance=5000

\begin{abstract}
In this work we shall study the late-time dynamics of several
$F(R)$ gravity models. By appropriately expressing the field
equations in terms of the redshift and of a statefinder function,
we shall solve numerically the field equations using appropriate
physical motivated initial conditions. We consider models which,
by construction, are described by a nearly $R^2$-model at early
epochs and we fine tune the parameters to achieve viability and
compatibility with the latest Planck constraints at late times.
Such models provide a unified description of inflation and dark
energy era and notably a common feature of all the models is the
presence of dark energy oscillations. Furthermore, we show that,
in contrast to general relativistic fluids and scalar field
descriptions, a large spectrum of different dark energy physics is
generated by simple $F(R)$ gravity models, varying from phantom,
to nearly de Sitter and to quintessential dark energy eras.
\end{abstract}

\maketitle

\section{Introduction}

Unambiguously, the observation that the Universe is currently
accelerating is one of the most unexpected and fascinating feature
of the current cosmological evolution. Dark energy is a mystery in
modern theoretical cosmology, and several theoretical proposals
can in principle describe successfully this cosmological era. In
the context of simple general relativity, the dark energy era can
be described by quintessence fields, and also by the rather
unmotivated and unappealing interacting fluids approach. However,
the scalar field description fails to describe the dark energy era
in a self-consistent way, if the dark energy era is phantom, which
is a probability according to the current Planck cosmological
constraints on cosmological parameters \cite{Planck:2018vyg}.
Modified gravity in its various forms
\cite{reviews1,reviews2,reviews3,reviews4,reviews5} can
successfully describe the dark energy era in a self-consistent
way, without resorting to phantom scalar fields, in order to have
an equation of state (EoS) parameter slightly tuned to the
phantom. The most prominent and simpler among modified gravities
is $F(R)$ gravity
\cite{Nojiri:2003ft,Capozziello:2005ku,Capozziello:2004vh,Capozziello:2018ddp,Hwang:2001pu,Cognola:2005de,Nojiri:2006gh,Song:2006ej,Capozziello:2008qc,Bean:2006up,Capozziello:2012ie,Faulkner:2006ub,Olmo:2006eh,Sawicki:2007tf,Faraoni:2007yn,Carloni:2007yv,
Nojiri:2007as,Capozziello:2007ms,Deruelle:2007pt,Appleby:2008tv,Dunsby:2010wg,Odintsov:2020nwm,Odintsov:2019mlf,Odintsov:2019evb,Oikonomou:2020oex,Oikonomou:2020qah},
in the context of which various cosmological eras can successfully
be realized. More importantly, in the context of $F(R)$ gravity
inflation and the dark energy era can be described in a unified
way, see the pioneer work \cite{Nojiri:2003ft}, and also
\cite{Nojiri:2006gh,Nojiri:2007as,Appleby:2008tv,Odintsov:2019evb,Oikonomou:2020oex,Oikonomou:2020qah}
for some recent developments. At first time the unification of the
inflation with dark energy epoch in $F(R)$ gravity was proposed in
\cite{Nojiri:2003ft}. After that, several realistic models of
$F(R)$ gravity unifying inflation with dark energy, and consistent
with observational bounds were developed in
\cite{Cognola:2007zu,Nojiri:2007cq}. In this line of research, in
this work we shall present several dark energy $F(R)$ gravity
models, which primordially are described by the $R^2$ model
\cite{Starobinsky:1980te,Bezrukov:2007ep} and at late-times all
these models mimic the $\Lambda$-Cold-Dark-Matter model
($\Lambda$CDM). All the models we shall present realize a viable
dark energy era, with dark energy EoS parameter and dark energy
density parameter compatible with the latest Planck constraints on
cosmological parameters. With this work we aim to present a
panorama of viable $f(R)$ gravity dark energy models which can
also generate a viable inflationary era.

This paper is organized as follows: In section II we overview the
formulation of $F(R)$ gravity which is suitable for describing the
late-time dynamics of $F(R)$ gravity. We shall express the field
equations in terms of the redshift as a dynamical parameter and we
shall use a suitable statefinder parameter, capable of containing
all the information needed for late-time dynamics. In section III
we present several viable $F(R)$ gravity models which can produce
a successful late-time era, in addition to a successful
inflationary era. Finally, the conclusions follow at the end of
the article.

\section{$F(R)$ Gravity Late-time Evolution Framework}

Consider an $F(R)$ gravity model in the presence of perfect matter
fluids, with action,
\begin{equation}
\label{action} \centering
\mathcal{S}=\int{d^4x\sqrt{-g}\left(\frac{F(R)}{2\kappa^2}+\mathcal{L}_m\right)}\,
,
\end{equation}
where $g$ denotes the determinant of the metric tensor $g^{\mu
\nu}$ and $\mathcal{L}_m$ stands for the Lagrangian of the perfect
matter fluids that are considered present. The term $F(R)$
describes an arbitrary function of the Ricci scalar $R$ and in our
case it will be taking the form of,
\begin{equation}\label{FR}
    F(R)=R+f(R) .
\end{equation}
We also note that $\kappa ^2=8 \pi G=\frac{1}{M_p^2}$, where $G$
is Newton's constant and $M_p$ is the reduced Planck mass. For the
background metric, we select the flat Friedmann-Robertson-Walker
(FRW) metric with the following line element,
\begin{equation}
    \centering\label{frw}
    d s^2 = - d t^2 + a(t) \sum_{i = 1}^3 d x_i^2,
\end{equation}
where $\alpha(t)$ is the scale factor. Given that the Ricci scalar
for the FRW metric is equal to,
\begin{equation}\label{ricciscalar}
    R = 6 \dot{H} + 12 H^2,
\end{equation}
by varying the action (\ref{action}) with respect to the metric,
we obtain the gravitational equations of motion,
\begin{equation} \label{Friedman}
    3 F_R H^2=\kappa^2 \rho_m + \frac{F_R R - F}{2} -3H \dot F_R \, ,
\end{equation}
\begin{equation} \label{Raycha}
    -2 F_R \dot H = \kappa^2 (\rho_m + R_m) + \ddot F -H \dot F \, ,
\end{equation}
where $F_R = \frac{\partial F}{\partial R}$ and the ``dot''
denotes derivative with respect to cosmic time. $H=\frac{\dot
\alpha}{\alpha}$ is the Hubble parameter and $\rho_m , P_m$ stand
for the matter fluids energy density and the corresponding
pressure respectively. We can actually rewrite the field equations
(\ref{Friedman}),(\ref{Raycha}) in the Einstein-Hilbert form for a
flat FRW metric and we get,
\begin{equation} \label{Friedtot}
    3 H^2= \kappa^2 \rho_{tot}  \ ,
\end{equation}
\begin{equation} \label{Raychtot}
   -2 \dot H =\kappa^2 (\rho_{tot} + P_{tot}) \ ,
\end{equation}
where $\rho_{tot}$ is the total energy density of the effective
cosmological fluid and $P_{tot}$ the corresponding pressure of it.
We assume that the cosmological fluid has 3 main contributions,
those of cold dark matter ($\rho_m$), of radiation ($\rho_r$) and
of dark energy ($\rho_{DE}$) . Therefore we have,
$\rho_{tot}=\rho_m + \rho_r + \rho_{DE}$ and $P_{tot}=P_m + P_r +
P_{DE}$. The late-time evolution is predominantly driven by the
dark energy fluid, the characteristics of which can be read off
the Friedmann and Raychaudhuri equations, (\ref{Friedman}) and
(\ref{Raycha}),
\begin{equation}\label{rDE}
    \rho_{DE}=\frac{F_R R - F}{2} + 3 H^2 (1-F_R)-3H \dot F_R \, ,
\end{equation}
\begin{equation}\label{PDE}
    P_{DE}=\ddot F -H \dot F +2 \dot H (F_R -1) - \rho_{DE} \, .
\end{equation}
However, instead of the cosmic time we prefer using the redshift as a dynamical variable to quantify evolution, which is defined as,
\begin{equation}
    1+z=\frac{1}{a} ,
\end{equation}
where we assumed that the present time scale factor is equal to
unity, therefore $z=0$ at present. Furthermore, we introduce the
statefinder function $y_H (z)$
\cite{Hu:2007nk,Bamba:2012qi,Odintsov:2020vjb,Odintsov:2020nwm,Odintsov:2020qyw,reviews1},
\begin{equation} \label{yhdef}
    y_H(z)=\frac{\rho_{DE}}{\rho_m^{(0)}}=\frac{H^2}{m_s^2}-(1+z)^3-\chi (1+z)^4 ,
\end{equation}
with $\rho_m^{(0)}$ denoting the energy density of the cold dark
matter at present time, $m_s^2=\frac{\kappa^2
\rho_m^{(0)}}{3}=H_0^2 \Omega_m=1.37 \times 10^{-67} eV^2$ is the
mass scale and $\chi$ is defined as
$\chi=\frac{\rho_r^{(0)}}{\rho_m^{(0)}} \simeq 3.1 \times
10^{-4}$, where $\rho_r^{(0)}$ is the present time radiation
energy density.

By combining the equations (\ref{Friedtot}) , (\ref{FR}) and
(\ref{yhdef}), we can express the Friedmann equation in terms of
the statefinder $y_H$ and it reads,
\begin{equation} \label{difyh}
    \frac{d^2 y_H}{dz^2} + J_1 \frac{d y_H}{dz} + J_2 y_H + J_3=0 \, ,
\end{equation}
where the dimensionless functions $J_1$ , $J_2$ , $J_3 $ are
defined as follows,
\begin{equation} \label{J1}
    J_1 = \frac{1}{(z+1)} \Big( -3-\frac{1}{y_H + (z+1)^3 + \chi (z+1)^4}  \frac{1-F_R}{6 m_s^2F_{RR}}\Big) \, ,
\end{equation}

\begin{equation} \label{J2}
    J_2 = \frac{1}{(z+1)^2} \Big( \frac{1}{y_H + (z+1)^3 +\chi (z+1)^4} \frac{2-F_R}{3 m_s^2 F_{RR}}\Big) \, ,
\end{equation}

\begin{equation} \label{J3}
    J_3 = -3(z+1) - \frac{(1-F_R)((z+1)^3 + 2\chi (z+1)^4) + (R-F)/(3 m_s^2)}{(z+1)^2 (y_H + (z+1)^3 +\chi (z+1)^4)} \frac{1}{6 m_s^2 F_{RR}}\, ,
\end{equation}
and $F_{RR}=\frac{\partial^2 F}{\partial R^2}$. Also, the Ricci
scalar as a function of the Hubble rate and of the redshift is
equal to,
\begin{equation}\label{ft10newadd}
R=12H^2-6HH_{z}(1+z)\, ,
\end{equation}
so the Ricci scalar is an implicit function of the statefinder
parameter $y_H$ and can be expressed in terms of it as follows,
\begin{equation}\label{neweqnrefricciyH}
R(z)=3\,m_s^2\left(-(z+1)\,\frac{d y_H(z)}{dz} + 4 y_H(z) +
(1+z)^3\right)\, .
\end{equation}
In order to study the late-time evolution of the universe using an
$F(R)$ gravity approach, we need to solve (\ref{difyh})
numerically for the redshift interval $z=[0,10]$ which demands
determining the initial conditions. Note that the derivatives of
the statefinder quantity $y_H(z)$ in terms of the redshift also
appear in the dimensionless parameters $J_i$, see Eq.
(\ref{neweqnrefricciyH}). Thus our aim is to solve numerically the
differential equation (\ref{difyh}), subject to physically
motivated initial conditions. We consider the following physically
motivated choice of initial conditions at $z_f=10$
\cite{Bamba:2012qi,Odintsov:2020vjb,Odintsov:2020nwm,Odintsov:2020qyw,reviews1},
\begin{equation}\label{initialcond}
    y_H (z_f) = \frac{ \Lambda}{3 m_s^2} \Big( 1 + \frac{1+z_f}{1000} \Big) \ , \ \frac{d y_H(z)}{dz} \Big |_{z=z_f} = \frac{1}{1000} \frac{ \Lambda}{3 m_s^2},
\end{equation}
where $\Lambda \simeq 11.895 \times 10^{-67} eV^2$.  The numerical
analysis will yield the statefinder quantity $y_H(z)$ as a
function of the redshift. From the obtained numerical solution
$y_H(z)$, one can evaluate all the relevant physical quantities,
such as the Hubble rate, the Ricci scalar, the energy density
parameter $\Omega_{DE}(z)$, the dark energy EoS parameter, the
total EoS parameter and the deceleration parameter. Specifically,
the Hubble rate in terms of $y_H(z)$ is,
\begin{equation}\label{hubblefr}
H(z)=m_s\sqrt{y_H(z)+(1+z)^{3}+\chi (1+z)^4}\, .
\end{equation}
where $\chi$ is defined below Eq. (\ref{yhdef}). Also the Ricci
scalar describing the curvature reads,
\begin{equation}\label{curvature}
    R(z)=3 m_s^2 \Big( 4 y_H(z) -(z+1) \frac{d y_H (z)}{dz} + (z+1)^3 \Big),
\end{equation}
and the energy density parameter $\Omega_{DE}(z)$ is given by,
\begin{equation}\label{OmegaDE}
    \Omega_{DE}(z)=\frac{y_H(z)}{y_H(z)+(z+1)^3 + \chi (z+1)^4}.
\end{equation}
The dark energy EoS parameter is,
\begin{equation}\label{EoSDE}
    \omega_{DE}(z)=-1+\frac{1}{3}(z+1)\frac{1}{y_H(z)}\frac{d y_H(z)}{dz},
\end{equation}
while the total EoS parameter reads,
\begin{equation}\label{EoStot}
    \omega_{tot}(z)=\frac{2(z+1)H'(z)}{3H(z)}-1 \ ,
\end{equation}
and finally the deceleration parameter is,
\begin{equation}\label{declpar}
    q(z)=-1-\frac{\dot H}{H^2}=-1-(z+1)\frac{H'(z)}{H(z)},
\end{equation}
where the ``prime'' denotes derivative with respect to the
redshift. We also note that the Hubble rate is given by Eq.
(\ref{hubblefr}), while for the $\Lambda CDM$ model equals to
\begin{equation}\label{hubblelcdm}
    H_{\Lambda}(z)=H_0\sqrt{\Omega_{\Lambda} +\Omega_M(z+1)^3 +\Omega_r(z+1)^4 } ,
\end{equation}
where $\Omega_{\Lambda} \simeq 0.68136$ and $\Omega_M \simeq
0.3153$. Also $ H_0 \simeq 1.37187 \times 10^{-33}$eV\footnote{The
conversion of the Hubble rate to the usual physical units
$\mathrm{km}/s\times \mathrm{Mpc}^{-1}$ can easily be done by
using 1sec=9.715$\times 10^{15}$eV$^{-1}$,
1Mpc=1.56373$\times10^{29}$eV$^{-1}$ and 1km=$5.06765\times
10^{9}$eV$^{-1}$.} is the present time Hubble rate according to
the latest Planck data \cite{Planck:2018vyg}.

Let us briefly discuss at this point the choice of the initial
conditions (\ref{initialcond}) and the motivation for choosing
these conditions. For a deeper analysis on this issue see also
\cite{Bamba:2012qi}. Basically the initial conditions
(\ref{initialcond}) correspond to the behavior of the statefinder
function $y_H(z)$ at large redshifts, so deeply in the matter
domination era, multibillion years in our Universe's past, well
before the deceleration to acceleration transition point.
Basically the redshift for which the initial conditions
(\ref{initialcond}) should be valid is possibly well before $z\sim
2$. In such a case, $y_H(z)\ll (1+z)^3$ so one may solve the
differential equation (\ref{difyh}) around the auxiliary redshift
$z_0$, which is a redshift well after the deceleration to
acceleration transition corresponding to the late-time era. Deeply
in the matter domination era, the scalar curvature is
approximately $R\sim 3m_s^2(1+z)^3$. So by considering the
differential equation (\ref{difyh}) around $z\sim z_0+(z-z_0)$
with $|z-z_0|\ll z$, and keeping leading order terms, the
differential equation (\ref{difyh}) becomes,
\begin{equation}\label{approximatediffeqn}
\frac{d^2 y_H}{dz^2} + \frac{\alpha_1}{z-z_0} \frac{d y_H}{dz} +
\frac{\beta_1}{(z-z_0)^2} y_H =\zeta_0+\zeta_1(z-z_0) \, ,
\end{equation}
where $\zeta_0$ and $\zeta_1$ are constants, and also $\alpha_1$
and $\beta_1$ are,
\begin{equation}\label{vetaandalpha}
\alpha_1=-\frac{7}{2}-\frac{(1-F_R(R_0))F_{RRR}(R_0)}{2F_{RR}(R_0)^2},\,\,\,\beta_1=2+\frac{1}{R_0F_{RR}(R_0)}+2\frac{(1-F_R(R_0))F_{RRR}(R_0)}{F_{RR}(R_0)^2}\,
,
\end{equation}
where $R_0$ is the curvature scalar at the redshift $z_0$. So by
solving the differential equation (\ref{approximatediffeqn}) one
obtains at leading order (we omit exponential decaying terms),
\begin{equation}\label{solution}
y_H(z)\simeq \alpha_1+\beta_1(z-z_0)\, .
\end{equation}
Hence at a redshift $z_f$ deeply in the matter domination era, one
has a linear dependence of the statefinder function $y_H(z)$ in
terms of the redshift. So this basically motivates the initial
conditions of Eq. (\ref{initialcond}), for which we also allowed a
physically motivated fine-tuning regarding the parameters
$\alpha_1$ and $\beta_1$. It is notable though that the overall
qualitative behavior of the models we shall present does not
change, however if the initial conditions are changed, the values
of the free parameters of the models that render each model
viable, change, as is expected.

Also it is vital that every dark energy model which we shall
discuss in the next section, satisfies the viability criteria that
every $F(R)$ gravity must satisfy. These are
\cite{reviews1,Zhao:2008bn},
\begin{equation}\label{viabilitycriteria}
F'(R)>0\,, \,\,\,F''(R)>0\, ,
\end{equation}
for $R>R_0$, with $R_0$ being the curvature at present day. So for
each model these criteria should be satisfied, for all the
curvatures from the present day up to the inflationary era, where
$R\sim 12H_I^2$, with $H_I$ being the inflationary scale, which is
assumed to be of the order $H_I\sim 10^{13}$GeV.

\section{Late-time $F(R)$ Gravity dynamics}

In this section we present the results of the numerical analysis
mentioned in the previous section for seven $F(R)$ gravity models.
In essence, we calculate the present time values and plot the
evolution of some of the aforementioned statefinder and physical
quantities in the redshift interval $z=[0,10]$ and compare them
with the $\Lambda CDM$ Model or the latest Planck data on
cosmological parameters \cite{Planck:2018vyg}. Another remark to
be made is that we added a $R^2/M^2$ term in every function, where
$M=3.04375 \times 10^{22}$eV. The value for the parameter $M$, its
actual value is determined by inflationary phenomenological
reasons, and it is equal to $M= 1.5\times
10^{-5}\left(\frac{N}{50}\right)^{-1}M_p$ \cite{Appleby:2009uf},
where $N$ being the $e$-foldings number during the inflationary
era. So for $N\sim \mathcal{O}(50-60)$ one gets the value
$M=3.04375 \times 10^{22}$eV. The $R^2$ term has a key role in the
unification of the early-time and the late-time eras, since it is
the dominant term in the evolution of the early universe, where
$R\sim H_I^2$ where $H_I$ is the inflationary scale., but becomes
insignificant compared to the other terms at late times where $R$
becomes comparable to the cosmological constant. On the contrary
when $R\sim H_I^2$ where $H_I$ is the inflationary scale, the
$R^2$ term dominates over the dark energy terms an thus the
inflationary era is controlled by the $R^2$ model.

\subsection{Type I $\Lambda$CDM-like with logarithmic $F(R)$ Gravity: Model 1}

\begin{figure}
\centering
\includegraphics[width=18pc]{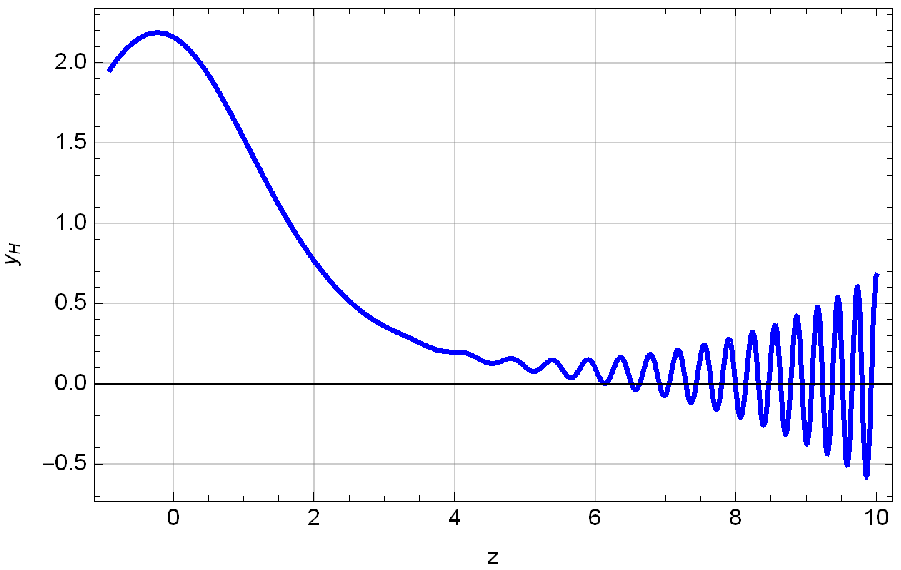}
\includegraphics[width=18pc]{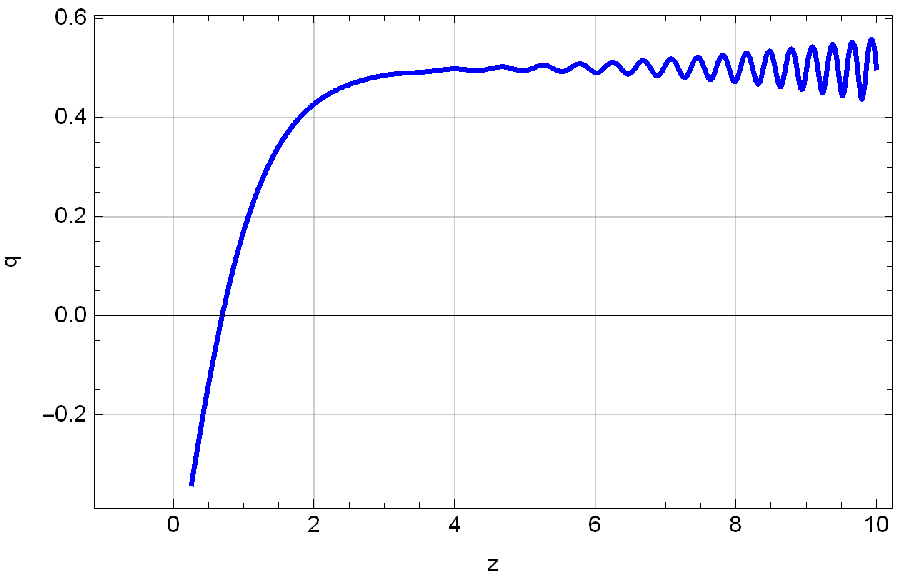}
\includegraphics[width=18pc]{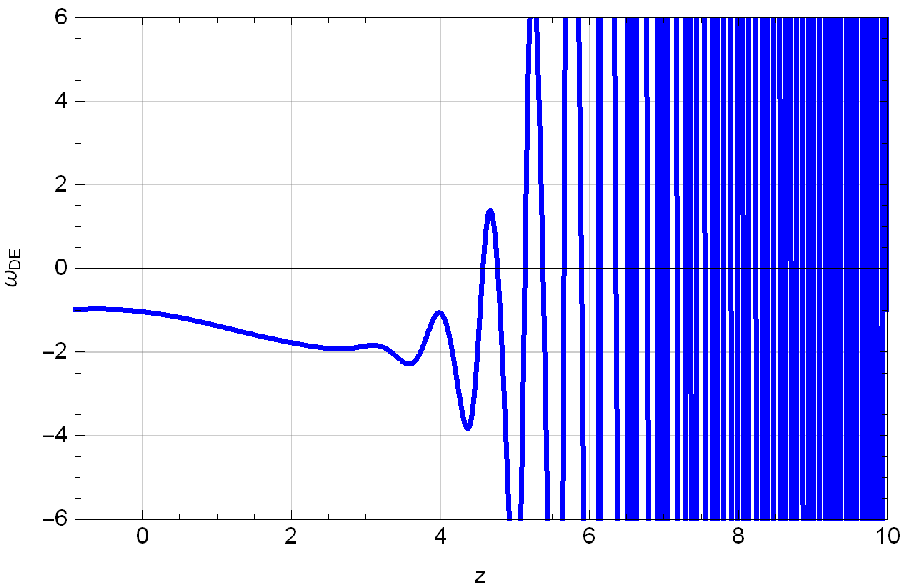}
\caption{Plots of the statefinder quantities $y_H(z)$ (upper left
plot), the deceleration parameter $q(z)$ (upper right plot) and
the dark energy EoS parameter $\omega_{DE}(z)$ (lower plot) as
functions of the redshift for the logarithmic type $F(R)$ model of
Eq. (\ref{fr20}) for $b=0.5 m_s^2$ and $R_0=220
m_s^2$.}\label{20pl}
\end{figure}

\begin{figure}
\centering
\includegraphics[width=18pc]{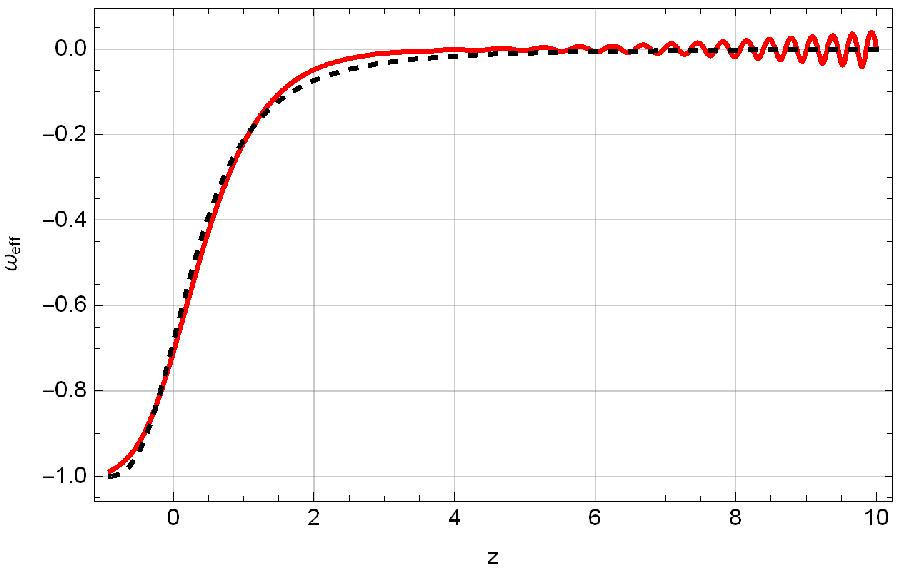}
\caption{Plot of total (effective) EoS parameter for the
logarithmic type $F(R)$ model of Eq. (\ref{fr20}) for $b=0.5
m_s^2$ and $R_0=220 m_s^2$ (red line) and for the $\Lambda$CDM
model (dashed line) as functions of the redshift. }\label{20pl2}
\end{figure}

We shall begin with the following $F(R)$ function,
\begin{equation}\label{fr20}
    F(R)=R+\frac{R^2}{M^2}+\frac{b}{1+\log{(R/R_0)}},
\end{equation}
where $\log$ denotes the base 10 logarithm and $b$ and $R_0$ are
free parameters with dimensions of $eV^2$ (dimensions of $[m]^2$
in natural units). We choose the values $b=0.5 m_s^2$ and $R_0=220
m_s^2$ and using Eqs. (\ref{difyh})-(\ref{hubblelcdm}) we find
that for $z=0$, that is the present time, the values of the dark
energy density parameter and EoS parameter are
$\Omega_{DE}(0)=0.6834$ and $\omega_{DE}(0)=-1.0372$, which fall
into the viability limits of the Planck constraints,
$\Omega_{DE}=0.6847 \pm 0.0073$ and $\omega_{DE}=-1.018 \pm
0.031$. In Fig.\ref{20pl} we plot $\omega_{DE}$ for the redshift
interval $z=[0,10]$ and we observe that the oscillations amplify
as the redshift increases. Unfortunately, oscillations are a
characteristic of the behavior of the statefinders in most $F(R)$
gravity theories in the redshift interval $z=[5,10]$, and this a
well known feature that has also been pointed out in the
literature, see
\cite{Hu:2007nk,Bamba:2012qi,Elizalde:2011ds,Nadkarni-Ghosh:2021mws}.
In Fig. \ref{20pl} we also plot the deceleration parameter, with
$q(0)=-0.5632$ and the function $y_H$. Furthermore, in Fig.
\ref{20pl2} we plot the total (effective) EoS parameter, which has
a present time value of $\omega_{tot}(0)=-0.7088$. This $F(R)$
Gravity model provides a viable late-time phenomenology which is
proved by the values of the parameters that either comply with the
latest constraints or are very close to the observed ones, as one
can see in Table. \ref{table1}.
One indication that this model can give good results is given by
the plots of Fig. \ref{20pl}. The deceleration parameter varies
from $q(0) \simeq -0.5$ to $q(10) \simeq 0.5$, illustrating the
passage from a decelerating era to an accelerated one.
Furthermore, the total EoS parameter is almost equal to the dark
energy EoS parameter, both being driven to $\simeq 1$ and
$\omega_{tot}(10) \simeq 0$, indicating that a (decelerating)
matter dominated era is followed by a (accelerating) dark energy
dominated era. The Hubble parameter is found equal to $H=1.36918
\times 10^{-33}$eV, a value very close to the observed one today
based on the CMB measurements.

\subsection{Type I $\Lambda$CDM-like with logarithmic format $F(R)$: Model 2}

\begin{figure}
\centering
\includegraphics[width=18pc]{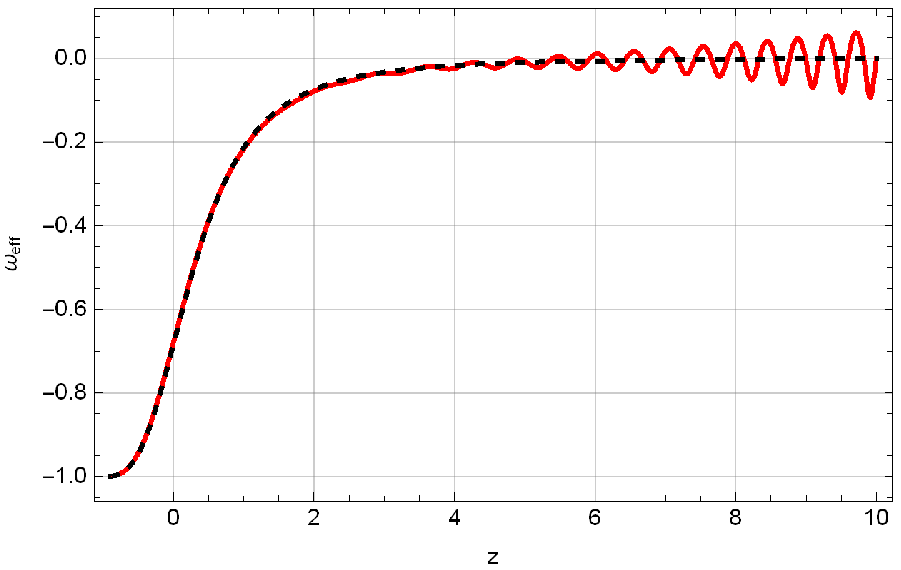}
\caption{Plot of total (effective) EoS parameter for the
logarithmic type $F(R)$ model of Eq. (\ref{fr22}) for
$b=11.81\Lambda, c=1.5 , R_0=m_s^2/100$ (red line) and for the
$\Lambda$CDM model (dashed line) as functions of the redshift.
}\label{22pl2}
\end{figure}

Moving on, we examine another logarithmic type $F(R)$ function,
\begin{equation}\label{fr22}
    F(R)=R+\frac{R^2}{M^2}-\frac{b}{c+1/\log(R/R_0)},
\end{equation}
where, as before, $b$ and $R_0$ are free parameters with
dimensions of $eV^2$ (dimensions of $[m]^2$ in natural units) and
$c$ is a dimensionless free parameter. After some investigation we
concluded that by setting the free parameters' values equal to
$b=11.81\Lambda, c=1.5 , R_0=m_s^2/100$ we get a viable
phenomenology. Specifically, the present time values for the dark
energy density parameter, the dark energy EoS parameter and the
total EoS parameter are $\Omega_{DE}(0)=0.6876$,
$\omega_{DE}(0)=-0.9891$ and $\omega_{tot}=-0.6801$ respectively.
The values of the first two parameters comply with the Planck
constraints and the latter is very close to the value calculated
for the $\Lambda$CDM model, that is $\omega_{\Lambda
tot}=-0.6847$. The similarity between the evolution of the total
EoS parameter between the two models is obvious in Fig.
\ref{22pl2}.
\begin{figure}
\centering
\includegraphics[width=18pc]{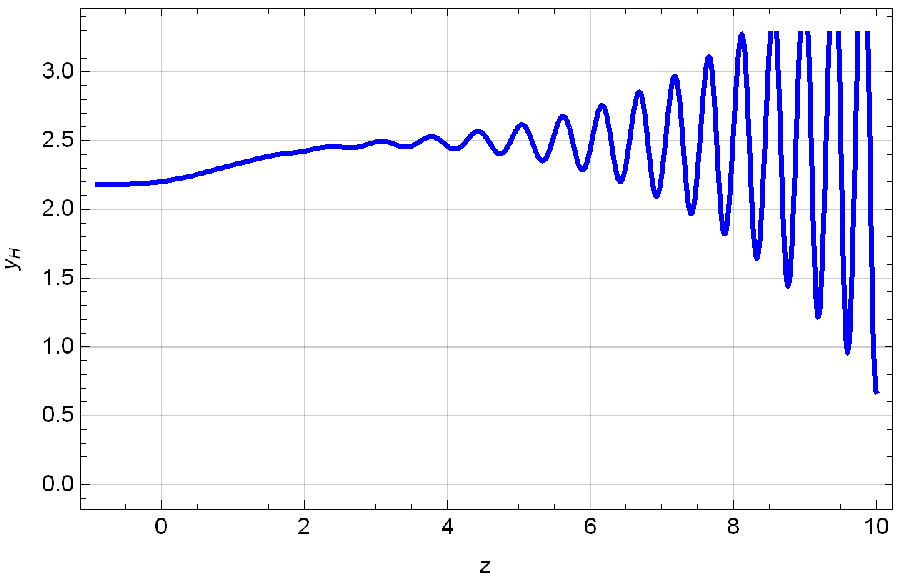}
\includegraphics[width=18pc]{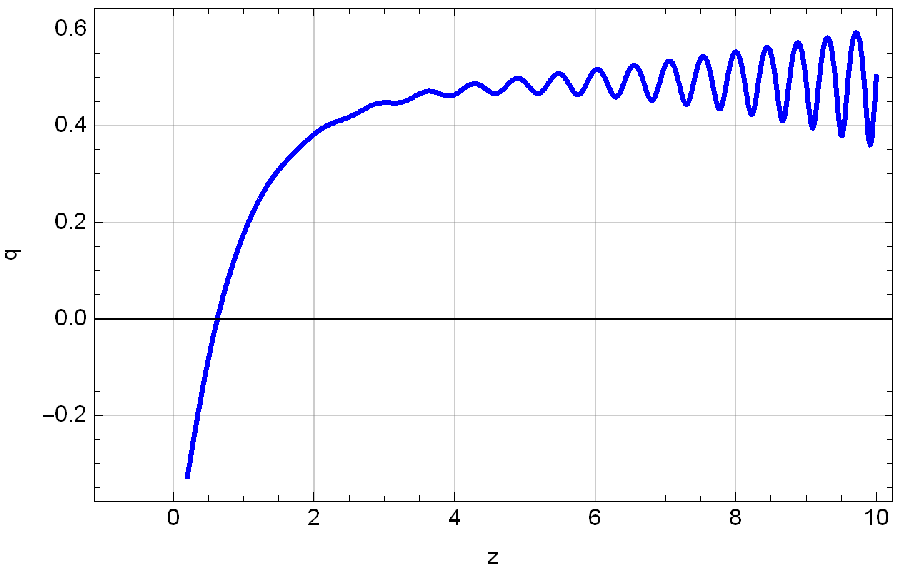}
\includegraphics[width=18pc]{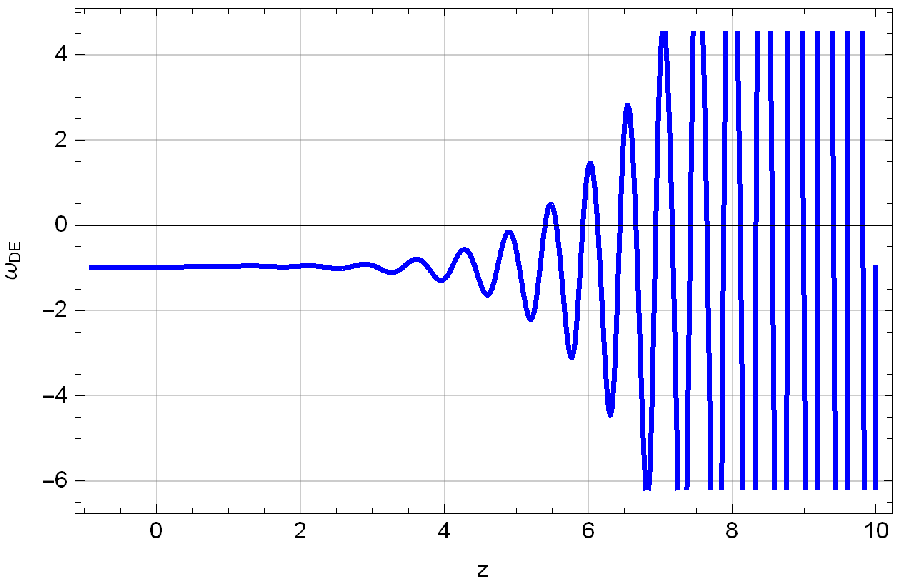}
\caption{Plots of the statefinder quantities $y_H(z)$ (upper left
plot), the deceleration parameter $q(z)$ (upper right plot) and
the dark energy EoS parameter $\omega_{DE}(z)$ (lower plot) as
functions of the redshift for the logarithmic type $F(R)$ model of
Eq. (\ref{fr22}) for $b=11.81\Lambda, c=1.5 ,
R_0=m_s^2/100$.}\label{22pl}
\end{figure}
In Fig. \ref{22pl} we also plot the functions $y_H(z)$ and $q(z)$,
with the latter taking the value $q(0)=-0.5201$, close to the one
from the $\Lambda$CDM too. Finally, we shall mention that the
calculated Hubble rate is $H=1.3785 \times 10^{-33}$eV. The
qualitative remarks from the plots are similar to the ones
mentioned for the previous case.

\subsection{Type I $\Lambda$CDM-like with exponential format $F(R)$ Model}

\begin{figure}
\centering
\includegraphics[width=18pc]{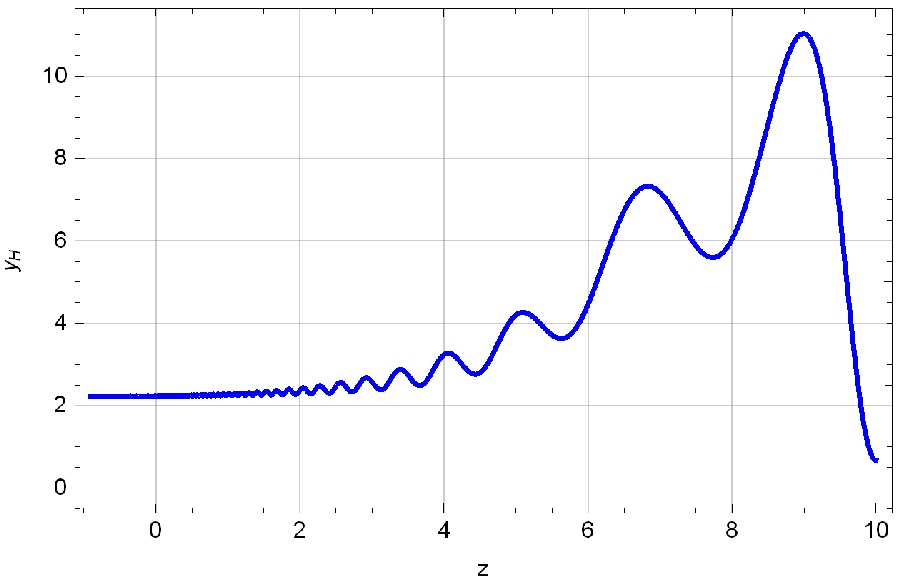}
\includegraphics[width=18pc]{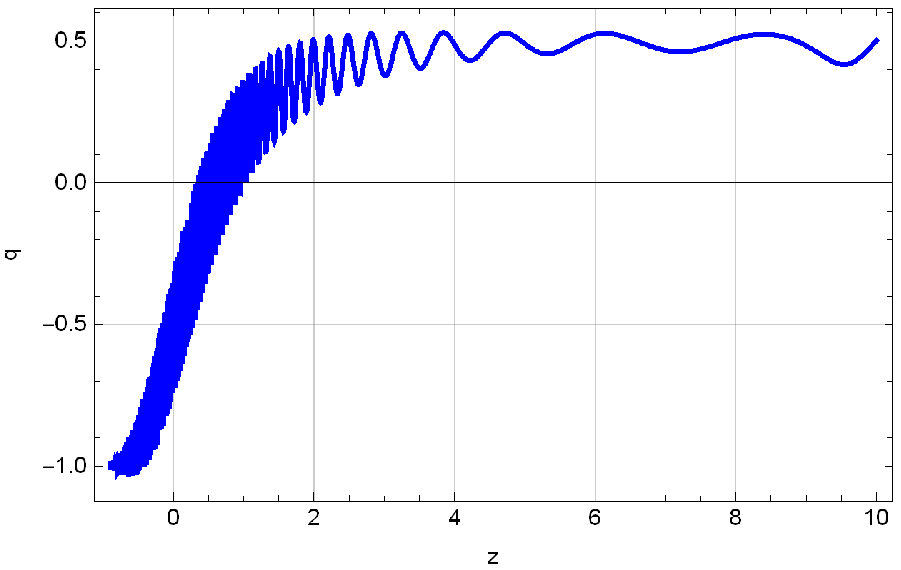}
\includegraphics[width=18pc]{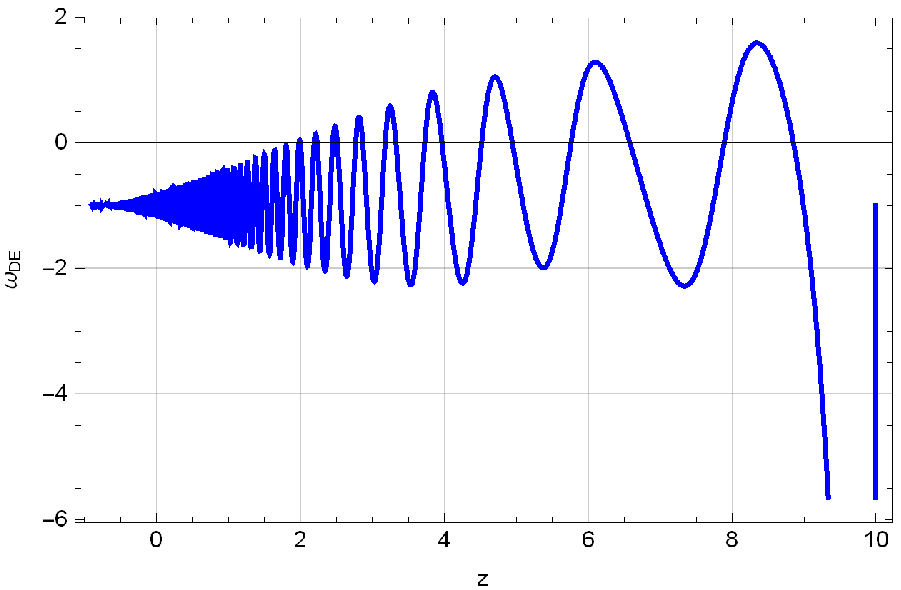}
\caption{Plots of the statefinder quantities $y_H(z)$ (upper left
plot), the deceleration parameter $q(z)$ (upper right plot) and
the dark energy EoS parameter $\omega_{DE}(z)$ (lower plot) as
functions of the redshift for the logarithmic type $F(R)$ model of
Eq. (\ref{fr24}) for $b=20\Lambda , c=2$ and
$R_0=m_s^2/0.00091$.}\label{24pl}
\end{figure}

\begin{figure}
\centering
\includegraphics[width=18pc]{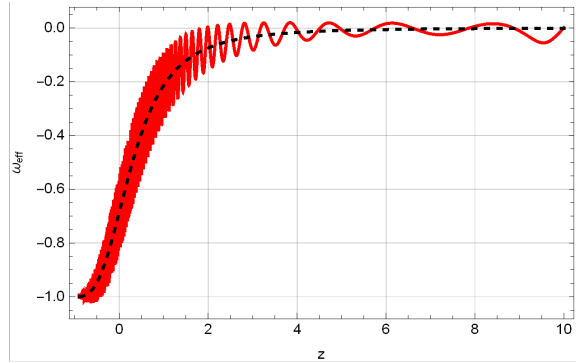}
\caption{Plot of total (effective) EoS parameter for the
logarithmic type $F(R)$ model of Eq. (\ref{fr24}) for
$b=20\Lambda, c=2$ and $R_0=m_s^2/0.00091$ (red line) and for the
$\Lambda$CDM model (dashed line) as functions of the redshift.
}\label{24pl2}
\end{figure}

Let us proceed to the results obtained from the study of the following $F(R)$ model,
\begin{equation}\label{fr24}
    F(R)=R+\frac{R^2}{M^2}-\frac{b}{c+\exp(-R/R_0)},
\end{equation}
where, as per usual, $b$ and $R_0$ are free parameters with
dimensions of $eV^2$ (dimensions of $[m]^2$ in natural units) and
$c$ is a dimensionless free parameter. We followed the numerical
solution path described in the previous section and by setting
$b=20\Lambda , c=2$ and $R_0=m_s^2/0.00091$ we concluded to a
viable dark energy for $F(R)$ Gravity model. The plot of $y_H(z)$
is given in Fig. \ref{24pl} and using this solution of
(\ref{difyh}) along with Eq. (\ref{hubblefr})-(\ref{EoStot}) we
can construct the rest of the plots of Fig. \ref{24pl} and Fig.
\ref{24pl2} and calculate the present day values of the
cosmological quantities of interest. The dark energy energy
density parameter and EoS parameter at $z=0$ are evaluated to be
$\Omega_{DE}=0.6918$ and $\omega_{DE}=-0.9974$, while the total
EoS parameter is found equal to $\omega_{tot}=-0.6899$. Moreover,
the deceleration parameter at present time is estimated equal to
$q(0)=-0.5349$ and the Hubble rate equal to $H=1.3878\times
10^{-33}$eV. The values of the cosmological parameters
$\Omega_{DE}$ and $\omega_{DE}$ comply with the latest Planck
constraints and, generally, the values of all the aforementioned
parameters belong within the observational regime.
One characteristic of this model is that the oscillations
accompanying the evolution of these cosmological parameters are
evident in the whole $z=[0,10]$ redshift range, losing amplitude
but rising in frequency as we move from $z=10$ to $z=0$. It is
widely suggested that the oscillations in $F(R)$ gravity theories
are a model-dependent characteristic, therefore we attribute this
to the exponential nature of the model.

\subsection{Type I $\Lambda$CDM-like with polynomial format $F(R)$ Model}

\begin{figure}
\centering
\includegraphics[width=18pc]{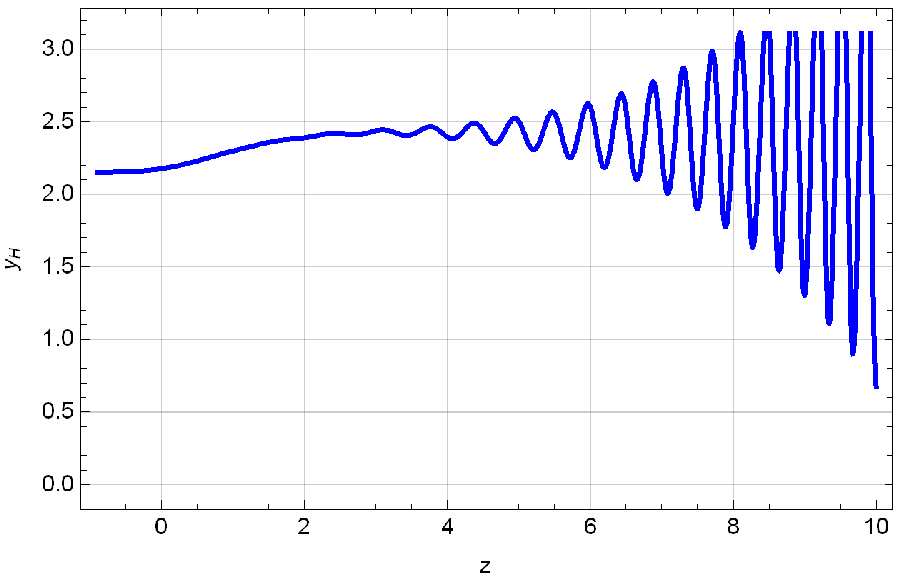}
\includegraphics[width=18pc]{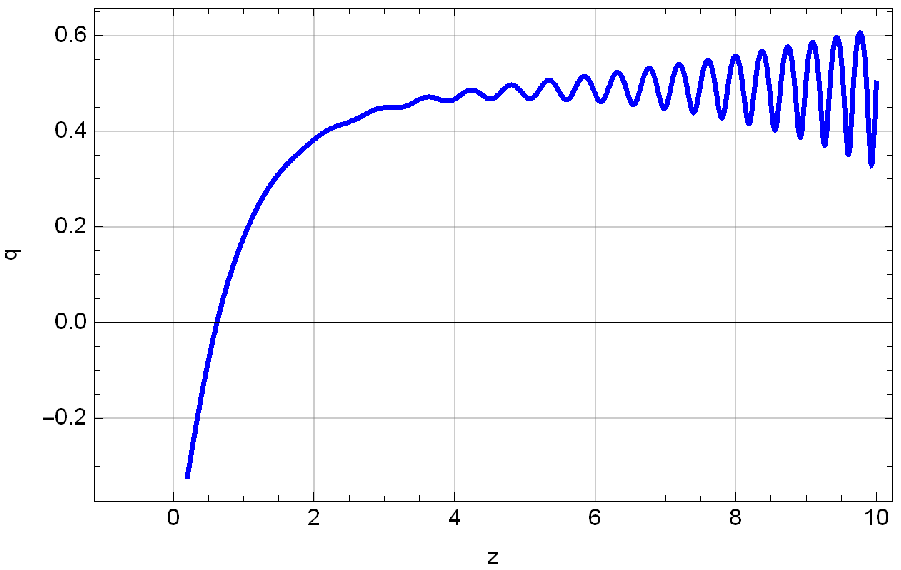}
\includegraphics[width=18pc]{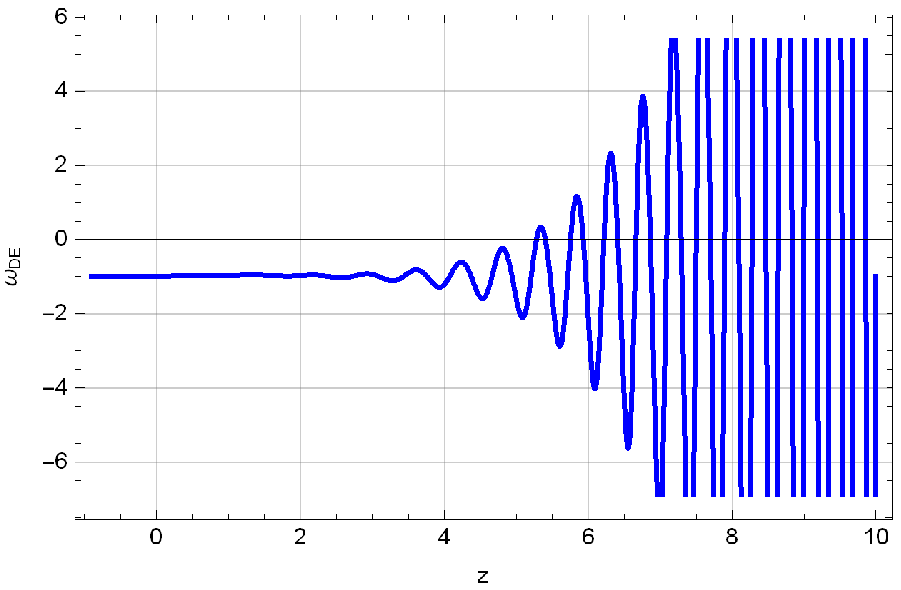}
\caption{Plots of the statefinder quantities $y_H(z)$ (upper left
plot), the deceleration parameter $q(z)$ (upper right plot) and
the dark energy EoS parameter $\omega_{DE}(z)$ (lower plot) as
functions of the redshift for the logarithmic type $F(R)$ model of
Eq. (\ref{frhu}) for $\alpha=1.4 \Lambda$, $b=1$, $c=0.2$ ,
$d=0.04$, $R_0=m_s^2$ and $n=0.3$.}\label{hupl}
\end{figure}

\begin{figure}
\centering
\includegraphics[width=18pc]{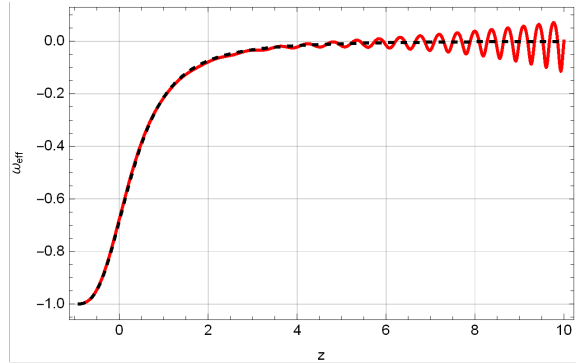}
\caption{Plot of total (effective) EoS parameter for the
logarithmic type $F(R)$ model of Eq. (\ref{frhu}) for $\alpha=1.4
\Lambda$, $b=1$, $c=0.2$ , $d=0.04$, $R_0=m_s^2$ and $n=0.3$ (red
line) and for the $\Lambda$CDM model (dashed line) as functions of
the redshift. }\label{hupl2}
\end{figure}

Another $F(R)$ model with interesting results comes from the following function,
\begin{equation}\label{frhu}
    F(R)=R+\frac{R^2}{M^2}-\alpha \frac{b(R/R_0)^n}{c(R/R_0)^n+d},
\end{equation}
where $b,c,d,n$ are dimensionless parameters and $\alpha,R_0$ are
free parameters with dimensions of $eV^2$ (dimensions of $[m]^2$
in natural units). We set $\alpha=1.4 \Lambda$, $b=1$, $c=0.2$ ,
$d=0.04$, $R_0=m_s^2$ and $n=0.3$ and following the numerical
analysis described in the previous section we find the present
time values of the statefinder quantities of interest. In Fig.
\ref{hupl} we plot the quantities $y_H$ , $q$, $\omega_{DE}$ and
in Fig. \ref{hupl2} the total EoS parameter $\omega_{tot}$ as
functions of the redshift for the interval $z=[0,10]$.
Specifically for the present day values, this model gives
$\Omega_{DE}(0)=0.6851$, $\omega_{DE}(0)=-0.9887$, $q(0)=-0.5160$,
$\omega_{tot}=-0.6773$ and $H=1.37295 \times 10^{-33}$eV. It is
obvious that for the chosen initial conditions and values for the
free parameters, the values of the cosmological quantities for
these model belong within the desired range.

\subsection{Type II $\Lambda$CDM-like with logarithmic format $F(R)$ model}

\begin{figure}
\centering
\includegraphics[width=18pc]{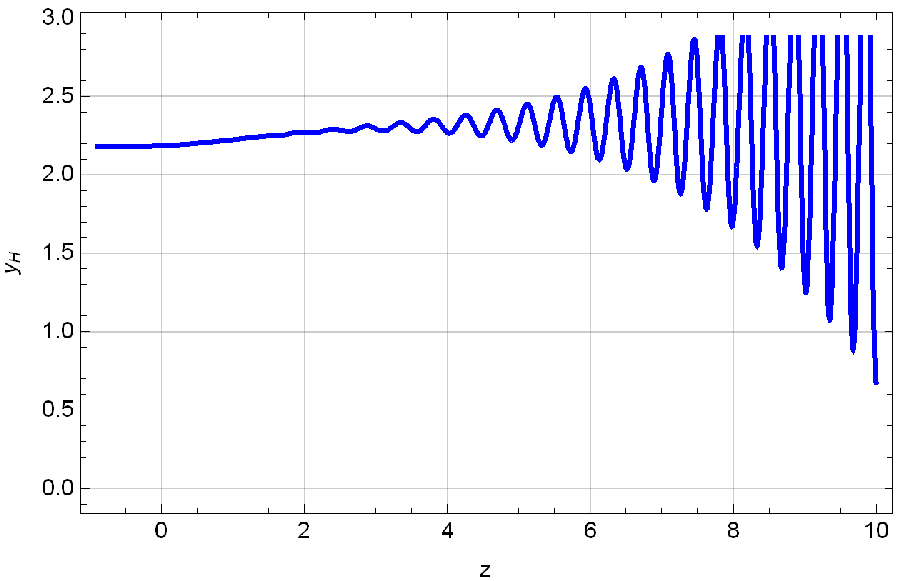}
\includegraphics[width=18pc]{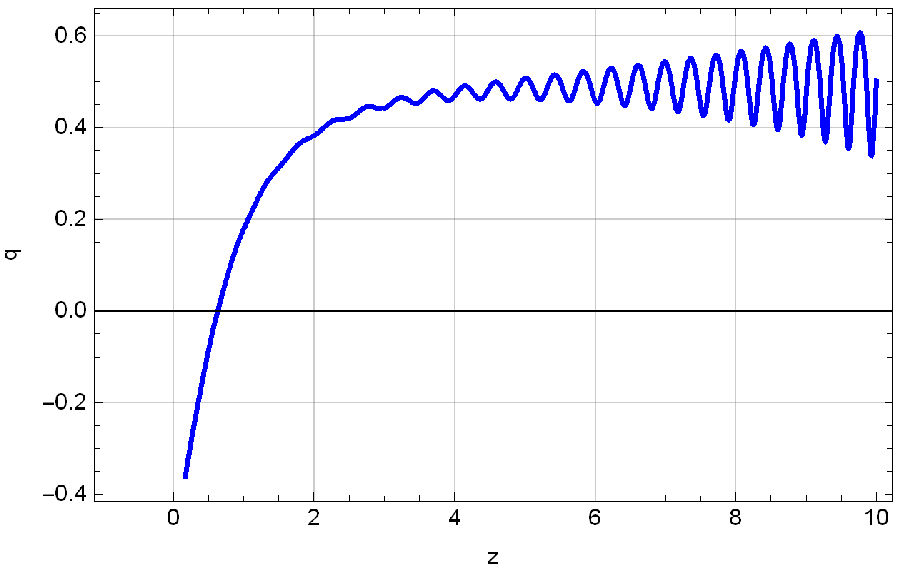}
\includegraphics[width=18pc]{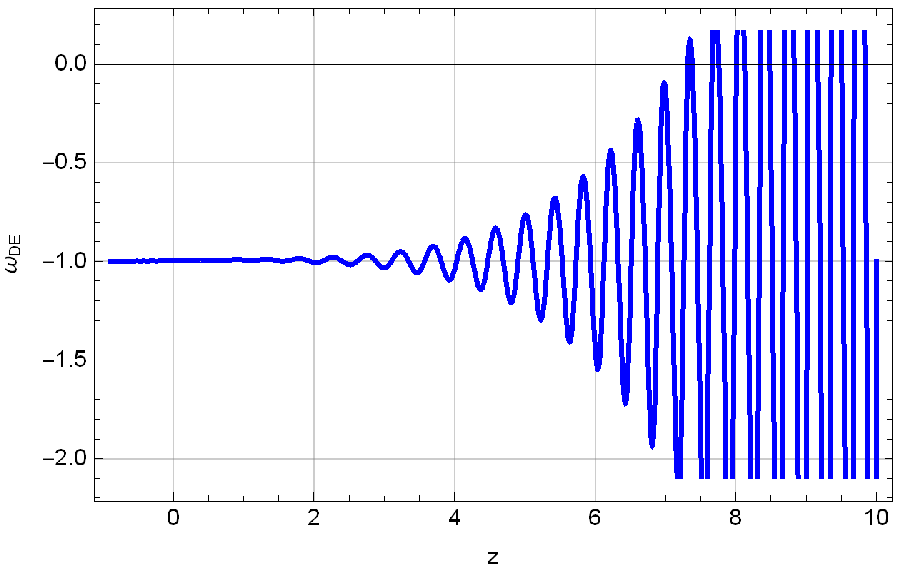}
\caption{Plots of the statefinder quantities $y_H(z)$ (upper left
plot), the deceleration parameter $q(z)$ (upper right plot) and
the dark energy EoS parameter $\omega_{DE}(z)$ (lower plot) as
functions of the redshift for the logarithmic type $F(R)$ model of
Eq. (\ref{fr29}) for $R_0=m_s^2 \times 10^{-64}$ and
$\alpha=10^{63} \Lambda/m_s^2=2.0045 \times 10^{63}$.}\label{29pl}
\end{figure}

\begin{figure}
\centering
\includegraphics[width=18pc]{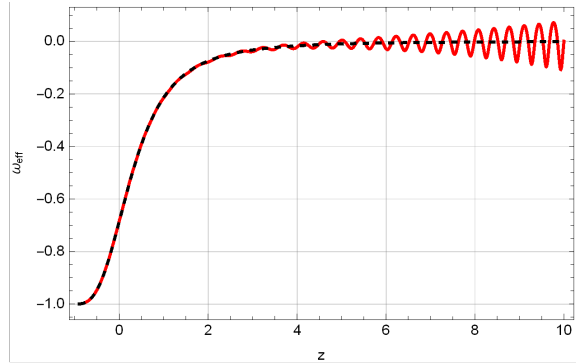}
\caption{Plot of total (effective) EoS parameter for the
logarithmic type $F(R)$ model of Eq. (\ref{fr29}) for $R_0=m_s^2
\times 10^{-64}$ and $\alpha=10^{63} \Lambda/m_s^2=2.0045 \times
10^{63}$ (red line) and for the $\Lambda$CDM model (dashed line)
as functions of the redshift.}\label{29pl2}
\end{figure}

Next we shall consider another type of $F(R)$ gravity function,
\begin{equation}\label{fr29}
    F(R)=R+\frac{R^2}{M^2}-\alpha R_0 \log(1+R/R_0),
\end{equation}
where $\alpha$ is a dimensionless free parameter and $R_0$ is
another free parameter with dimensions of $eV^2$ (dimensions of
$[m]^2$ in natural units). This type of $F(R)$ gravities were
introduced and studied in \cite{Nojiri:2003ni,Bamba:2014mua}. The
results from the study of this model for $R_0=m_s^2 \times
10^{-64}$ and $\alpha=10^{63} \Lambda/m_s^2=2.0045 \times 10^{63}$
are presented in Fig. \ref{29pl}, Fig. \ref{29pl2} and Table
\ref{table1}. In particular, for this $F(R)$ model we find that
$\Omega_{DE}(0)=0.68596$, which is compatible with the Planck
constraint $\Omega_{DE}=0.6847 \pm 0.0073$ and
$\omega_{DE}(0)=-0.9962$, which also is compatible with the Planck
constraint $\omega_{DE}=-1.018 \pm 0.031$. Additionally, we
evaluate the present time values of to total EoS parameter
$\omega_{tot}(0)=-0.6832$ , the deceleration parameter
$q(0)=-0.5249$ and the Hubble rate $H=1.37483 \times 10^{-33}$eV.
Thus, we have another model with viable dark energy phenomenology
for the values for the free parameters mentioned above and the
initial conditions (\ref{initialcond}).

\subsection{Type II $\Lambda$CDM-like with exponential format $F(R)$ model}

\begin{figure}
\centering
\includegraphics[width=18pc]{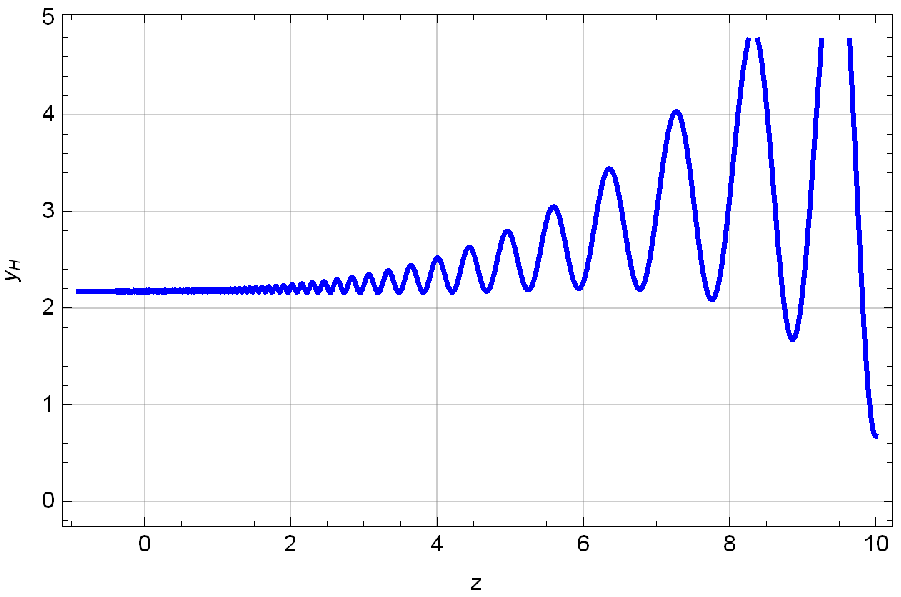}
\includegraphics[width=18pc]{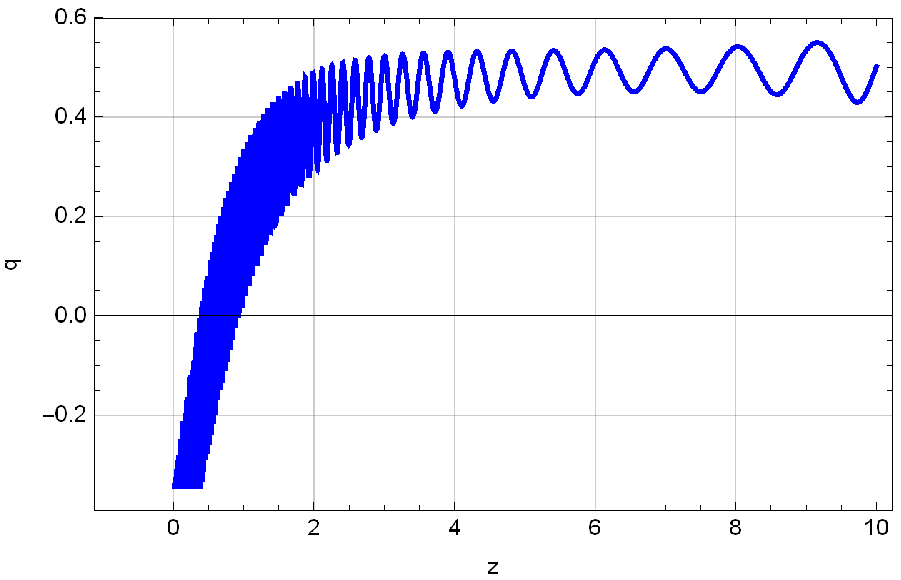}
\includegraphics[width=18pc]{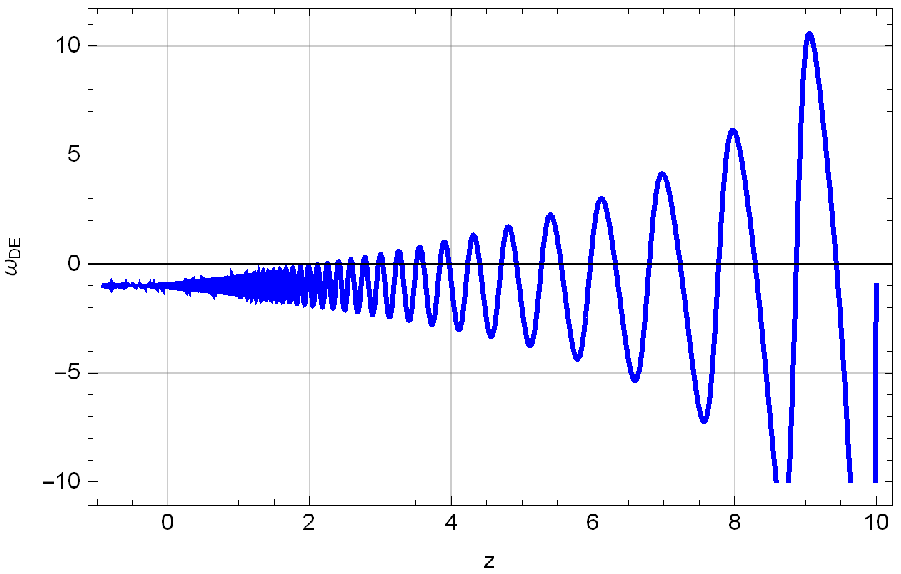}
\caption{Plots of the statefinder quantities $y_H(z)$ (upper left
plot), the deceleration parameter $q(z)$ (upper right plot) and
the dark energy EoS parameter $\omega_{DE}(z)$ (lower plot) as
functions of the redshift for the logarithmic type $F(R)$ model of
Eq. (\ref{fr32}) for $c=7.5$, $b=\Lambda$ and
$R_0=m_s^2/0.0005$.}\label{32pl}
\end{figure}

\begin{figure}
\centering
\includegraphics[width=18pc]{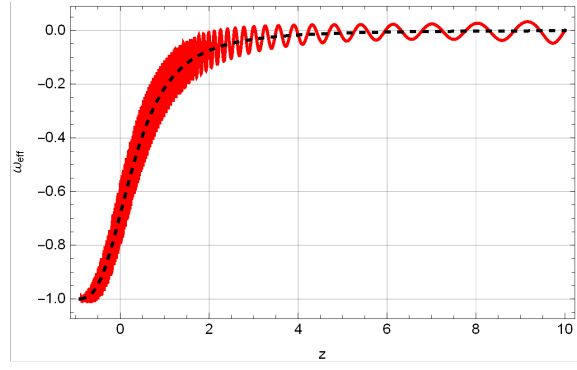}
\caption{Plot of total (effective) EoS parameter for the
logarithmic type $F(R)$ model of Eq. (\ref{fr32}) for $c=7.5$,
$b=\Lambda$ and $R_0=m_s^2/0.0005$ (red line) and for the
$\Lambda$CDM model (dashed line) as functions of the redshift.
}\label{32pl2}
\end{figure}

We shall proceed our examination with yet another exponential type $F(R)$ function,
\begin{equation}\label{fr32}
    F(R)=R+\frac{R^2}{M^2}-b[c-\exp(-R/R_0)],
\end{equation}
where $b,R_0$ are free parameters with dimensions of $eV^2$
(dimensions of $[m]^2$ in natural units) and $c$ is a
dimensionless free parameter. For these parameters we pick the
values $c=7.5$, $b=\Lambda$ and $R_0=m_s^2/0.0005$ and the
viability of this model is evident from the results shown in Fig.
\ref{32pl}, Fig. \ref{32pl2} and Table \ref{table1}. In detail,
for the dark energy density parameter we obtain
$\Omega_{DE}(0)=0.6847$ and for the dark energy EoS parameter
$\omega_{DE}(0)=-1.0367$, values that belong within the range of
the Planck constraints. As for the rest cosmological parameters we
find that $q(0)=-0.5647$, $\omega_{tot}= -0.7098$ and $H=1.37212
\times 10^{-33}$eV. At this point we shall discuss the qualitative
characteristics of the plots in Fig. \ref{32pl} and \ref{32pl2}.
It is obvious from Eqs. (\ref{hubblefr})-(\ref{EoStot}) that the
evolution of the cosmological parameters depends on the one of the
statefinder quantity $y_H$ and as can be seen in Fig. \ref{32pl} ,
$y_H$ is plagued with oscillations, as usual, with their amplitude
falling and their frequency rising significantly as we move
forward from the redshift $z=10$ to $z=0$, with that
characteristic being passed to the other parameters as well. This
observation was made before for another exponential-type $F(R)$
model, indicating that the nature of the models is indeed the
cause.


\subsection{Type III $\Lambda$CDM-like with polynomial format $F(R)$ model}

\begin{figure}
\centering
\includegraphics[width=18pc]{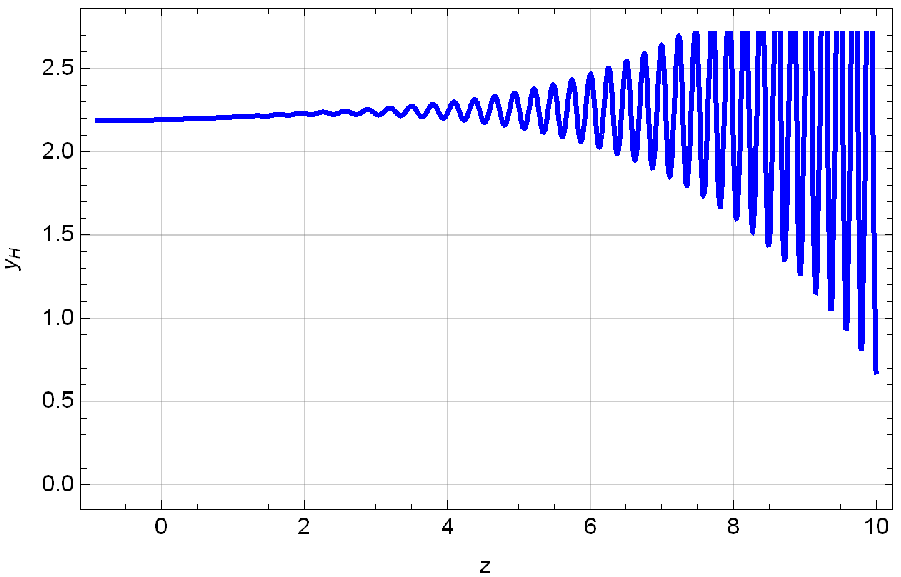}
\includegraphics[width=18pc]{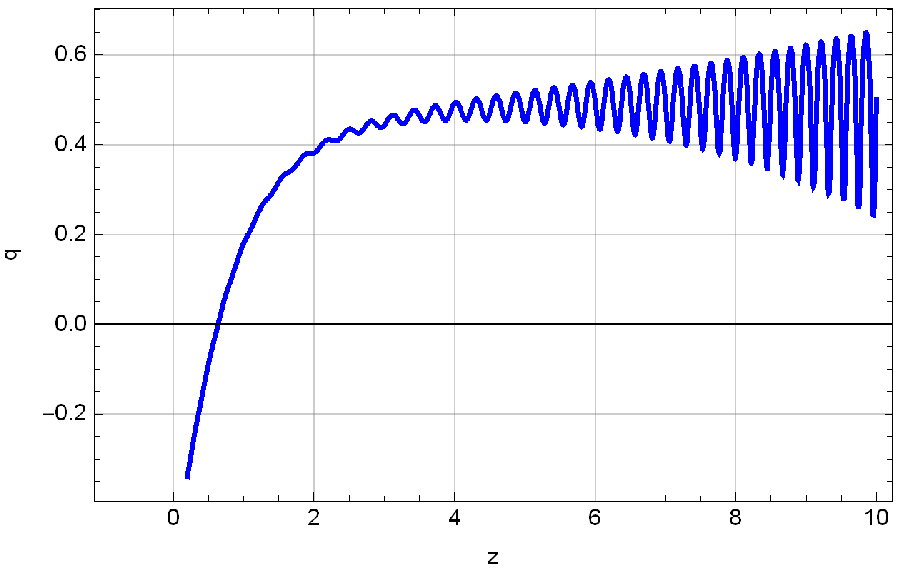}
\includegraphics[width=18pc]{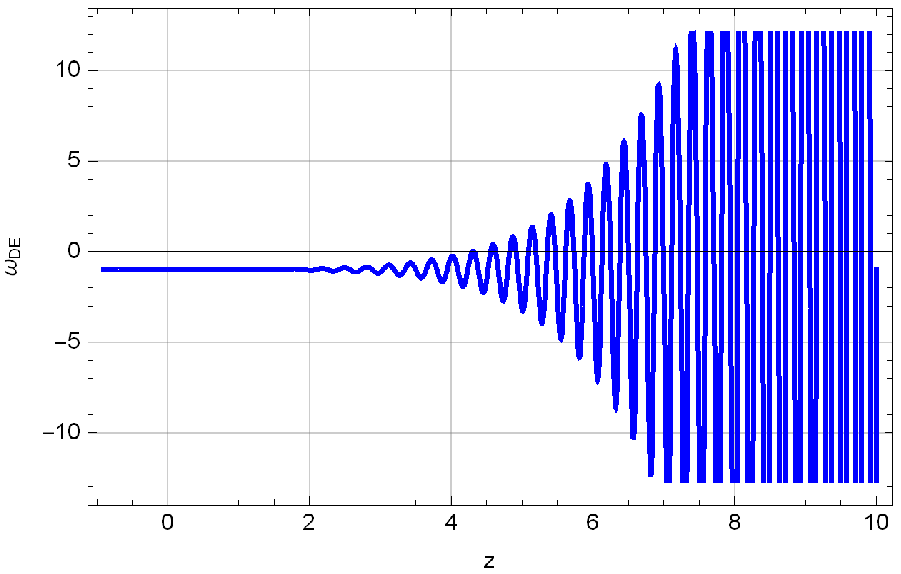}
\caption{Plots of the statefinder quantities $y_H(z)$ (upper left
plot), the deceleration parameter $q(z)$ (upper right plot) and
the dark energy EoS parameter $\omega_{DE}(z)$ (lower plot) as
functions of the redshift for the logarithmic type $F(R)$ model of
Eq. (\ref{fr37}) for  $b=\Lambda$ , $R_0=m_s^2$ , $c=27$, $d=3$
and $n=0.01$.}\label{37pl}
\end{figure}

\begin{figure}
\centering
\includegraphics[width=18pc]{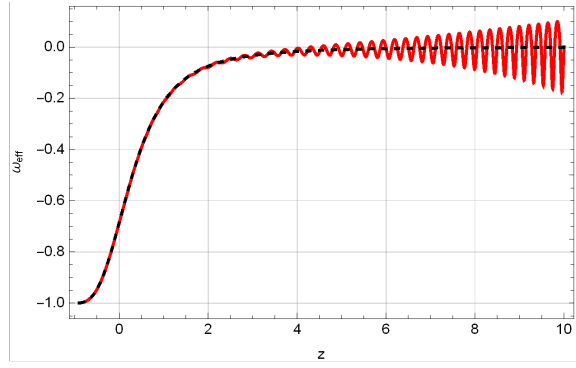}
\caption{Plot of total (effective) EoS parameter for the
logarithmic type $F(R)$ model of Eq. (\ref{fr37}) for $b=\Lambda$
, $R_0=m_s^2$ , $c=27$, $d=3$ and $n=0.01$ (red line) and for the
$\Lambda$CDM model (dashed line) as functions of the redshift.
}\label{37pl2}
\end{figure}

The last $F(R)$ function we shall study is,
\begin{equation}\label{fr37}
    F(R)=R+\frac{R^2}{M^2}-b\frac{c-(R_0/R)^n}{d+(R_0/R)^n},
\end{equation}
where $c,d,n$ are dimensionless free parameters and $b,R_0$ are
free parameters with dimensions of $eV^2$ (dimensions of $[m]^2$
in natural units). For this model, by setting $b=\Lambda$ ,
$R_0=m_s^2$, $c=27$ , $d=3$ and $n=0.01$ we find that
$\Omega_{DE}(0)=0.6867$ and $\omega_{DE}(0)=-0.9976$, which
satisfy the latest Planck mission constraints. Moreover, the
numerical analysis gives $\omega_{tot}=-0.6849$, $q(0)=-0.5274$
and $H=1.37626 \times 10^{-33}$eV, values very close to the
observed ones at present time. These results are also shown in
Table \ref{table1}, compared to the ones of the other models along
and the $\Lambda$CDM model or Planck constraints if provided. In
Fig. \ref{37pl} and Fig. \ref{37pl2} we plot the evolution of some
of the cosmological quantities for the redshift interval
$z=[0,10]$.
\begin{table}[h!]
  \begin{center}
    \caption{\emph{\textbf{Cosmological Parameters Values at present day for the $F(R)$
Gravity models studied in this work, quoting only their
corresponding Equation in the text, the $\Lambda$CDM Model and the
Planck 2018 data where available (quoting only the Planck 2018
constraint if available). The values for the $\Lambda$CDM model
are obtained by using the Hubble rate in Eq. (\ref{hubblelcdm})
and Eqs. (\ref{OmegaDE}), (\ref{EoSDE}), (\ref{EoStot}) and
(\ref{declpar}).}}}
    \label{table1}
    \begin{tabular}{|c|c|c|c|c|c|c|c|c|c|}
    \hline
      \textbf{Parameter} & \textbf{(\ref{fr20})}& \textbf{(\ref{fr22})} & \textbf{(\ref{fr24})}  & \textbf{(\ref{frhu})} & \textbf{(\ref{fr29})} & \textbf{(\ref{fr32})} & \textbf{(\ref{fr37})}& \makecell{\textbf{Planck 2018} \\ \textbf{or SNe I$A^a$ }}& \textbf{$\Lambda$CDM}
      \\  \hline
      $\Omega_{DE}(0)$ & $0.6834$ & $0.6876$ & $0.6918$ & $0.6851$ & $0.68596$ & $0.6847$ & $0.6867$ & $0.6847\pm 0.0073$&-

 \\  \hline
      $\omega_{DE}(0)$ & $-1.0372$ & $-0.9891$ & $-0.9974$ & $-0.9887$ & $-0.9962$ & $-1.0367 $ & $-0.9976$ & $-1.018\pm0.031$ & -

      \\  \hline
      $q(0)$ & $-0.5632$ & $-0.5201$ & $-0.5349$ & $-0.5160$ & $-0.5249$ & $-0.5647$ & $-0.5274$ & $-0.38 \pm 0.05$ (SNe I$A^a$) & $-0.535$

      \\  \hline
      $\omega_{tot}(0)$ & $-0.7088$ & $-0.6801$ & $-0.6899$ & $-0.6773$ & $-0.6832$ & $-0.7098$ & $-0.6849$ & - & $-0.68467$

      \\  \hline
      $H \times 10^{-33}$eV &  $1.36918$ & $1.3785$& $1.3878$ & $1.37295$ & $1.37483$ & $1.37212$ & $1.37626$ & $1.37187$ & $ 1.37187$

      \\  \hline
    \end{tabular}
  \end{center}
\end{table}

\begin{figure}[h!]
\centering
\includegraphics[width=20pc]{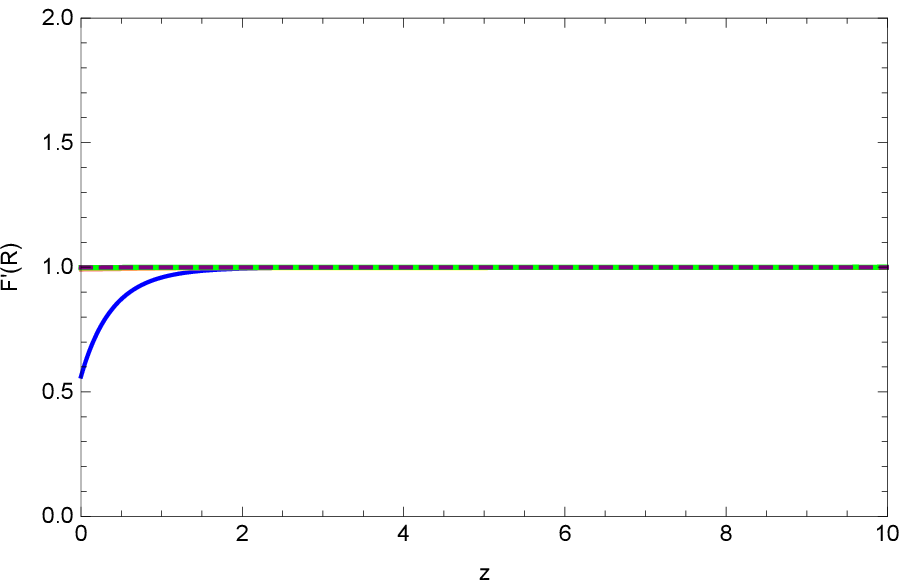}
\includegraphics[width=22pc]{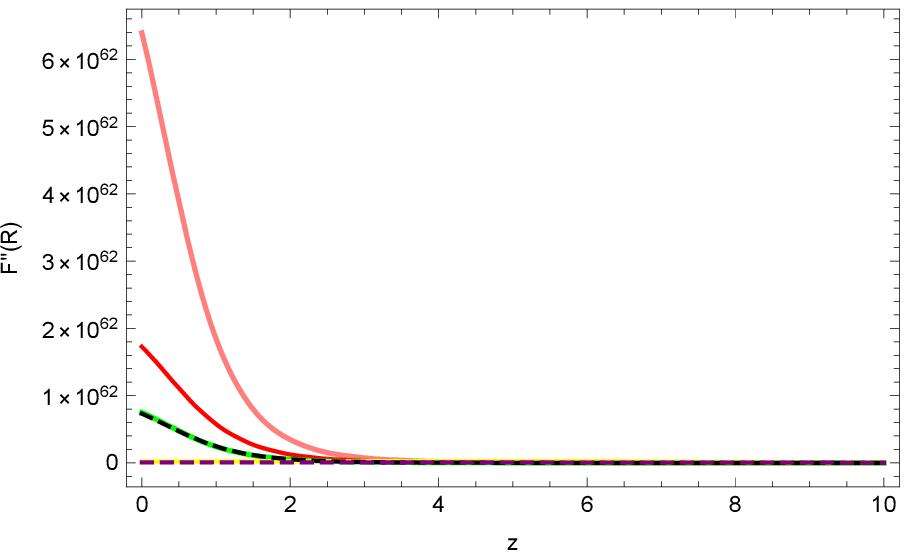}
\includegraphics[width=20pc]{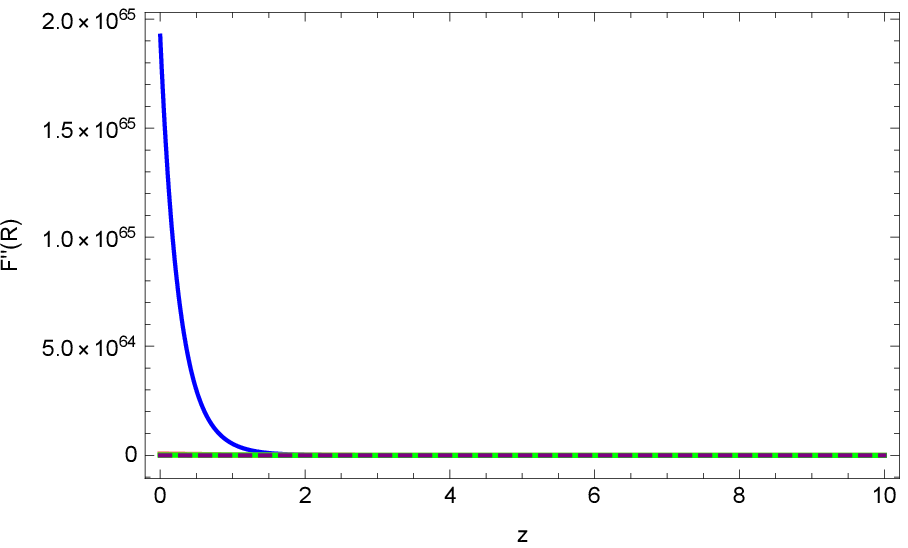}
\caption{The terms $F'(R)$ (upper left plot) and $F''(R)$ (upper
right and bottom plot) as functions of the redshift in the
redshift range $z=[0,10]$, for the models of Eqs. (\ref{fr20})
(blue curve), (\ref{fr22}) (pink curve), (\ref{fr24}) (yellow
curve), (\ref{frhu}) (green curve), (\ref{fr29}) (red curve),
(\ref{fr32}) (black curve) and (\ref{fr37}) (purple curve). The
viability criteria are satisfied for all the models.}
\label{viability1}
\end{figure}

At this point, let us investigate if the viability criteria
(\ref{viabilitycriteria}) are satisfied for all the models studied
in the this and the previous subsections. In Fig. \ref{viability1}
we plot the behavior of the derivatives $F'(R)$ (upper left plot)
and $F''(R)$ (upper right and bottom plots) in terms of the
redshift in the range $z=[0,10]$. The Model 1 of Eq. (\ref{fr20})
corresponds to the blue curve, the Model 2 of Eq. (\ref{fr22})
corresponds to the pink curve, the type I $\Lambda$CDM-like with
exponential format $F(R)$ model of Eq. (\ref{fr24}) corresponds to
the yellow curve, the type I $\Lambda$CDM-like with polynomial
format $F(R)$ model of Eq. (\ref{frhu}) corresponds to the green
curve, the type II $\Lambda$CDM-like with logarithmic format
$F(R)$ model of Eq. (\ref{fr29}) corresponds to the red curve, the
type III $\Lambda$CDM-like with polynomial format $F(R)$ model of
Eq. (\ref{fr32}) corresponds to the black dashed curve, and
finally the type II $\Lambda$CDM-like with exponential format
$F(R)$ model of Eq. (\ref{fr37}) corresponds to the purple dashed
curve. In the upper left plot of Fig. \ref{viability1}, where the
$F'(R)$ is plotted as a function of the redshift, it can be seen
that the all the models satisfy the viability criterion and also
that the models are identical similar regarding the values of
$F'(R)$. In fact, as it can be seen from the upper left plot, all
the models are indistinguishable. In the upper right plot of Fig.
\ref{viability1} we present the values of $F''(R)$ as a function
of the redshift for the models of Eqs. (\ref{fr22}), (\ref{fr24}),
(\ref{frhu}), (\ref{fr29}), (\ref{fr32}) and (\ref{fr37}), and it
can be seen that the viability criteria are satisfied and also
some of the models are indistinguishable. The model of  Eq.
(\ref{fr20}) is compared to the other models regarding the value
of $F''(R)$ in the bottom plot of Fig. \ref{viability1}, since it
is quite different from the other models, but regardless it also
satisfies the viability criteria.
The viability are also satisfied for higher redshifts, let us see
this explicitly by considering inflationary scales of the
curvature, when $H\sim H_I$, then $R\sim 12H_I^2$. For all the
models we have $F''(R)\sim 2.1588\times 10^{-45}$eV$^{-1}$, and
$F'(R)\sim 3.59056$, hence the viability conditions
(\ref{viabilitycriteria}) are satisfied too for inflationary
scales. Also it is important to note that the reason for which the
values of $F'(R)$ and $F''(R)$ are the same for all the models is
that the $F(R)$ gravity at early times is dominated by the $R^2$
term, and it can be checked that if we calculate $F'(R)$ and
$F''(R)$ by only keeping the $R^2$ term, we end up to the same
values for $F'(R)$ and $F''(R)$ we quoted above. This result was
expected, since the models were engineered in such a way so that
at early times the $R^2$ term dominates.

As an overall comment let us note that in principle there is a
large range of values of the free parameters of the models we
studied, for which the viability is guaranteed. However, the free
parameters depend strongly on the values of the rest of free
variables, so certain amount of fine-tuning is required so that
the viability of each model is obtained.

It is also notable to mention that the starting redshift $z_f$
used in the initial conditions (\ref{initialcond}) also affects
the viability of the models. We checked all the models by using
different values for the starting redshift and a general
conclusion is that as the initial redshift changes, the values of
the free parameters must be changed for each model in order to
achieve the viability of the models. This also mentioned in the
relevant literature \cite{Nadkarni-Ghosh:2021mws}. However we omit
the details for different starting redshifts for brevity, since
the overall qualitative picture is the same.

Finally, let us note that all the different seven models we
considered, have more or less the same late-time qualitative
behavior, since all of them mimic the $\Lambda$CDM model and
predict viable dark energy related physical parameters. The
differences is that some models predict a phantom evolution
(Models (25) and (30)) while others predict a quintessential
accelerating era (Models (26)-(29) and (31)).

\section{Conclusions}

In this paper we investigated several models of $F(R)$ gravity
which may generated a viable dark energy era at late times and an
$R^2$-like inflationary era primordially. We considered several
different functional forms for the $F(R)$ gravities, varying from
logarithmic models to exponential and power-law models. We
expressed the field equations as functions of the redshift and we
quantified our analysis by introducing a statefinder function
$y_H(z)$ which basically measures the deformation from the
Einstein-Hilbert gravity caused by the $F(R)$ gravity terms. By
appropriately choosing the initial conditions at the last stages
of the matter domination era for the statefinder function $y_H(z)$
and its derivative with respect to the redshift, we solved
numerically the field equations and we derived the behavior of
several physical quantities and statefinders. Specifically we were
interested in the total EoS parameter, the dark energy EoS
parameter, the dark energy density parameter and the deceleration
parameter, which is the most characteristic statefinder available.
As we demonstrated, a large spectrum of distinct late time
cosmologies can be realized by different $F(R)$ gravity models.
Specifically, it is possible to realize phantom, nearly de-Sitter
and quintessential dark energy eras with simple $F(R)$ gravity
models, without resorting to phantom scalar fields or relativistic
fluids of any sort. With this work we aimed to provide a panorama
of viable $F(R)$ gravity models which can describe inflation and
dark energy in an unified manner, and we also aimed to demonstrate
the large spectrum of different physics generated by simple $F(R)$
gravity models. It is a challenge to test some of these models
regarding their primordial gravitational waves predictions, the
reason being the behavior during the radiation domination era. In
a future work we aim to present the predictions for the primordial
gravitational waves spectrum of some of these models but also to
consider ways to reduce the dark energy oscillations, which are
present in all the models.


\begin{thebibliography}{99}



%\cite{Planck:2018vyg}
\bibitem{Planck:2018vyg}
N.~Aghanim \textit{et al.} [Planck],
%``Planck 2018 results. VI. Cosmological parameters,''
Astron. Astrophys. \textbf{641} (2020), A6 [erratum: Astron.
Astrophys. \textbf{652} (2021), C4]
doi:10.1051/0004-6361/201833910 [arXiv:1807.06209 [astro-ph.CO]].
%6801 citations counted in INSPIRE as of 23 Jan 2022



\bibitem{reviews1}
 S.~Nojiri, S.~D.~Odintsov and V.~K.~Oikonomou,
  %``Modified Gravity Theories on a Nutshell: Inflation, Bounce and Late-time Evolution,''
  Phys.\ Rept.\  {\bf 692} (2017) 1
  %doi:10.1016/j.physrep.2017.06.001
  [arXiv:1705.11098 [gr-qc]].
  %%CITATION = %doi:10.1016/j.physrep.2017.06.001;%%
  %21 citations counted in INSPIRE as of 20 Aug 2017

\bibitem{reviews2}


 S. Capozziello, M. De Laurentis,
   %``Extended Theories of Gravity,''
   Phys.\ Rept.\  {\bf 509}, 167 (2011);\\
   %[arXiv:1108.6266 [gr-qc]].
   %%CITATION = ARXIV:1108.6266;%%
 V.~Faraoni and S.~Capozziello,
  %``Beyond Einstein Gravity : A Survey of Gravitational Theories for Cosmology and Astrophysics,''
  Fundam.\ Theor.\ Phys.\  {\bf 170} (2010).
  %%doi:10.1007/978-94-007-0165-6
  %%CITATION = %doi:10.1007/978-94-007-0165-6;%%
  %64 citations counted in INSPIRE as of 16 Sep 2017



\bibitem{reviews3}
S. Nojiri, S.D. Odintsov,
  %``Introduction to modified gravity and gravitational alternative for dark
  %energy,''
  eConf {\bf C0602061}, 06 (2006)
  [Int.\ J.\ Geom.\ Meth.\ Mod.\ Phys.\  {\bf 4}, 115 (2007)].
  %[arXiv:hep-th/0601213];
  %%CITATION = 00436,4,115;%%


   \bibitem{reviews4}

S. Nojiri, S.D. Odintsov,
   %``Unified cosmic history in modified gravity: from F(R) theory to
   %Lorentz non-invariant models,''
   Phys.\ Rept.\  {\bf 505}, 59 (2011);
   %[arXiv:1011.0544 [gr-qc]].
   %%CITATION = ARXIV:1011.0544;%%




\bibitem{reviews5}

A.~de la Cruz-Dombriz and D.~Saez-Gomez,
  %``Black holes, cosmological solutions, future singularities, and their thermodynamical properties in modified gravity theories,''
  Entropy {\bf 14} (2012) 1717
  %doi:10.3390/e14091717
  [arXiv:1207.2663 [gr-qc]].
  %%CITATION = %doi:10.3390/e14091717;%%
  %125 citations counted in INSPIRE as of 15 Nov 2017



%%%%%%%%%%%%%%%%%%%%%%%%%%%%%%%%%%%%%%%%%%%%%%%%%%%%%%%%%%%%%%%%%%%%%%%%%%%%%%%%%%%%%%%%%fr



%\cite{Nojiri:2003ft}
\bibitem{Nojiri:2003ft}
S.~Nojiri and S.~D.~Odintsov,
%``Modified gravity with negative and positive powers of the curvature: Unification of the inflation and of the cosmic acceleration,''
Phys. Rev. D \textbf{68} (2003), 123512
doi:10.1103/PhysRevD.68.123512 [arXiv:hep-th/0307288 [hep-th]].
%1693 citations counted in INSPIRE as of 26 Oct 2021


%\cite{Capozziello:2005ku}
\bibitem{Capozziello:2005ku}
S.~Capozziello, V.~F.~Cardone and A.~Troisi,
%``Reconciling dark energy models with f(R) theories,''
Phys. Rev. D \textbf{71} (2005), 043503
doi:10.1103/PhysRevD.71.043503 [arXiv:astro-ph/0501426
[astro-ph]].
%409 citations counted in INSPIRE as of 26 Oct 2021



%\cite{Capozziello:2018ddp}
\bibitem{Capozziello:2018ddp}
S.~Capozziello, C.~A.~Mantica and L.~G.~Molinari,
%``Cosmological perfect-fluids in f(R) gravity,''
Int. J. Geom. Meth. Mod. Phys. \textbf{16} (2018) no.01, 1950008
doi:10.1142/S0219887819500087 [arXiv:1810.03204 [gr-qc]].
%80 citations counted in INSPIRE as of 26 Oct 2021



%\cite{Capozziello:2004vh}
\bibitem{Capozziello:2004vh}
S.~Capozziello, V.~F.~Cardone and M.~Francaviglia,
%``f(R) Theories of gravity in Palatini approach matched with observations,''
Gen. Rel. Grav. \textbf{38} (2006), 711-734
doi:10.1007/s10714-006-0261-x [arXiv:astro-ph/0410135 [astro-ph]].
%81 citations counted in INSPIRE as of 26 Oct 2021
















%\cite{Hwang:2001pu}
\bibitem{Hwang:2001pu}
J.~c.~Hwang and H.~Noh,
%``f(R) gravity theory and CMBR constraints,''
Phys. Lett. B \textbf{506} (2001), 13-19
doi:10.1016/S0370-2693(01)00404-X [arXiv:astro-ph/0102423
[astro-ph]].
%87 citations counted in INSPIRE as of 26 Oct 2021



%\cite{Cognola:2005de}
\bibitem{Cognola:2005de}
G.~Cognola, E.~Elizalde, S.~Nojiri, S.~D.~Odintsov and S.~Zerbini,
%``One-loop f(R) gravity in de Sitter universe,''
JCAP \textbf{02} (2005), 010 doi:10.1088/1475-7516/2005/02/010
[arXiv:hep-th/0501096 [hep-th]].
%384 citations counted in INSPIRE as of 26 Oct 2021



%\cite{Nojiri:2006gh}
\bibitem{Nojiri:2006gh}
S.~Nojiri and S.~D.~Odintsov,
%``Modified f(R) gravity consistent with realistic cosmology: From matter dominated epoch to dark energy universe,''
Phys. Rev. D \textbf{74} (2006), 086005
doi:10.1103/PhysRevD.74.086005 [arXiv:hep-th/0608008 [hep-th]].
%775 citations counted in INSPIRE as of 26 Oct 2021



%\cite{Song:2006ej}
\bibitem{Song:2006ej}
Y.~S.~Song, W.~Hu and I.~Sawicki,
%``The Large Scale Structure of f(R) Gravity,''
Phys. Rev. D \textbf{75} (2007), 044004
doi:10.1103/PhysRevD.75.044004 [arXiv:astro-ph/0610532
[astro-ph]].
%469 citations counted in INSPIRE as of 26 Oct 2021




%\cite{Capozziello:2008qc}
\bibitem{Capozziello:2008qc}
S.~Capozziello, V.~F.~Cardone and V.~Salzano,
%``Cosmography of f(R) gravity,''
Phys. Rev. D \textbf{78} (2008), 063504
doi:10.1103/PhysRevD.78.063504 [arXiv:0802.1583 [astro-ph]].
%131 citations counted in INSPIRE as of 26 Oct 2021







%\cite{Bean:2006up}
\bibitem{Bean:2006up}
R.~Bean, D.~Bernat, L.~Pogosian, A.~Silvestri and M.~Trodden,
%``Dynamics of Linear Perturbations in f(R) Gravity,''
Phys. Rev. D \textbf{75} (2007), 064020
doi:10.1103/PhysRevD.75.064020 [arXiv:astro-ph/0611321
[astro-ph]].
%301 citations counted in INSPIRE as of 26 Oct 2021






%\cite{Capozziello:2012ie}
\bibitem{Capozziello:2012ie}
S.~Capozziello and M.~De Laurentis,
%``The dark matter problem from f(R) gravity viewpoint,''
Annalen Phys. \textbf{524} (2012), 545-578
doi:10.1002/andp.201200109
%82 citations counted in INSPIRE as of 26 Oct 2021



%\cite{Faulkner:2006ub}
\bibitem{Faulkner:2006ub}
T.~Faulkner, M.~Tegmark, E.~F.~Bunn and Y.~Mao,
%``Constraining f(R) Gravity as a Scalar Tensor Theory,''
Phys. Rev. D \textbf{76} (2007), 063505
doi:10.1103/PhysRevD.76.063505 [arXiv:astro-ph/0612569
[astro-ph]].
%345 citations counted in INSPIRE as of 26 Oct 2021


%\cite{Olmo:2006eh}
\bibitem{Olmo:2006eh}
G.~J.~Olmo,
%``Limit to general relativity in f(R) theories of gravity,''
Phys. Rev. D \textbf{75} (2007), 023511
doi:10.1103/PhysRevD.75.023511 [arXiv:gr-qc/0612047 [gr-qc]].
%220 citations counted in INSPIRE as of 26 Oct 2021



%\cite{Sawicki:2007tf}
\bibitem{Sawicki:2007tf}
I.~Sawicki and W.~Hu,
%``Stability of Cosmological Solution in f(R) Models of Gravity,''
Phys. Rev. D \textbf{75} (2007), 127502
doi:10.1103/PhysRevD.75.127502 [arXiv:astro-ph/0702278
[astro-ph]].
%209 citations counted in INSPIRE as of 26 Oct 2021


%\cite{Faraoni:2007yn}
\bibitem{Faraoni:2007yn}
V.~Faraoni,
%``de Sitter space and the equivalence between f(R) and scalar-tensor gravity,''
Phys. Rev. D \textbf{75} (2007), 067302
doi:10.1103/PhysRevD.75.067302 [arXiv:gr-qc/0703044 [gr-qc]].
%165 citations counted in INSPIRE as of 26 Oct 2021



%\cite{Carloni:2007yv}
\bibitem{Carloni:2007yv}
S.~Carloni, P.~K.~S.~Dunsby and A.~Troisi,
%``The Evolution of density perturbations in f(R) gravity,''
Phys. Rev. D \textbf{77} (2008), 024024
doi:10.1103/PhysRevD.77.024024 [arXiv:0707.0106 [gr-qc]].
%147 citations counted in INSPIRE as of 26 Oct 2021



%\cite{Nojiri:2007as}
\bibitem{Nojiri:2007as}
S.~Nojiri and S.~D.~Odintsov,
%``Unifying inflation with LambdaCDM epoch in modified f(R) gravity consistent with Solar System tests,''
Phys. Lett. B \textbf{657} (2007), 238-245
doi:10.1016/j.physletb.2007.10.027 [arXiv:0707.1941 [hep-th]].
%392 citations counted in INSPIRE as of 26 Oct 2021



%\cite{Capozziello:2007ms}
\bibitem{Capozziello:2007ms}
S.~Capozziello, A.~Stabile and A.~Troisi,
%``The Newtonian Limit of f(R) gravity,''
Phys. Rev. D \textbf{76} (2007), 104019
doi:10.1103/PhysRevD.76.104019 [arXiv:0708.0723 [gr-qc]].
%192 citations counted in INSPIRE as of 26 Oct 2021


%\cite{Deruelle:2007pt}
\bibitem{Deruelle:2007pt}
N.~Deruelle, M.~Sasaki and Y.~Sendouda,
%``Junction conditions in f(R) theories of gravity,''
Prog. Theor. Phys. \textbf{119} (2008), 237-251
doi:10.1143/PTP.119.237 [arXiv:0711.1150 [gr-qc]].
%102 citations counted in INSPIRE as of 26 Oct 2021


%\cite{Appleby:2008tv}
\bibitem{Appleby:2008tv}
S.~A.~Appleby and R.~A.~Battye,
%``Aspects of cosmological expansion in F(R) gravity models,''
JCAP \textbf{05} (2008), 019 doi:10.1088/1475-7516/2008/05/019
[arXiv:0803.1081 [astro-ph]].
%104 citations counted in INSPIRE as of 26 Oct 2021






%\cite{Dunsby:2010wg}
\bibitem{Dunsby:2010wg}
P.~K.~S.~Dunsby, E.~Elizalde, R.~Goswami, S.~Odintsov and
D.~S.~Gomez,
%``On the LCDM Universe in f(R) gravity,''
Phys. Rev. D \textbf{82} (2010), 023519
doi:10.1103/PhysRevD.82.023519 [arXiv:1005.2205 [gr-qc]].
%122 citations counted in INSPIRE as of 26 Oct 2021


%\cite{Odintsov:2020nwm}
\bibitem{Odintsov:2020nwm}
S.~D.~Odintsov and V.~K.~Oikonomou,
%``Geometric Inflation and Dark Energy with Axion $F(R)$ Gravity,''
Phys. Rev. D \textbf{101} (2020) no.4, 044009
doi:10.1103/PhysRevD.101.044009 [arXiv:2001.06830 [gr-qc]].
%39 citations counted in INSPIRE as of 26 Oct 2021



%\cite{Odintsov:2019mlf}
\bibitem{Odintsov:2019mlf}
S.~D.~Odintsov and V.~K.~Oikonomou,
%``$f(R)$ Gravity Inflation with String-Corrected Axion Dark Matter,''
Phys. Rev. D \textbf{99} (2019) no.6, 064049
doi:10.1103/PhysRevD.99.064049 [arXiv:1901.05363 [gr-qc]].
%50 citations counted in INSPIRE as of 26 Oct 2021





%\cite{Odintsov:2019evb}
\bibitem{Odintsov:2019evb}
S.~D.~Odintsov and V.~K.~Oikonomou,
%``Unification of Inflation with Dark Energy in $f(R)$ Gravity and Axion Dark Matter,''
Phys. Rev. D \textbf{99} (2019) no.10, 104070
doi:10.1103/PhysRevD.99.104070 [arXiv:1905.03496 [gr-qc]].
%39 citations counted in INSPIRE as of 26 Oct 2021


%\cite{Oikonomou:2020oex}
\bibitem{Oikonomou:2020oex}
V.~K.~Oikonomou,
%``Rescaled Einstein-Hilbert Gravity from $f(R)$ Gravity: Inflation, Dark Energy and the Swampland Criteria,''
Phys. Rev. D \textbf{103} (2021) no.12, 124028
doi:10.1103/PhysRevD.103.124028 [arXiv:2012.01312 [gr-qc]].
%11 citations counted in INSPIRE as of 26 Oct 2021



%\cite{Oikonomou:2020qah}
\bibitem{Oikonomou:2020qah}
V.~K.~Oikonomou,
%``Unifying inflation with early and late dark energy epochs in axion $F(R)$ gravity,''
Phys. Rev. D \textbf{103} (2021) no.4, 044036
doi:10.1103/PhysRevD.103.044036 [arXiv:2012.00586 [astro-ph.CO]].
%16 citations counted in INSPIRE as of 26 Oct 2021


%\cite{Cognola:2007zu}
\bibitem{Cognola:2007zu}
G.~Cognola, E.~Elizalde, S.~Nojiri, S.~D.~Odintsov, L.~Sebastiani
and S.~Zerbini,
%``A Class of viable modified f(R) gravities describing inflation and the onset of accelerated expansion,''
Phys. Rev. D \textbf{77} (2008), 046009
doi:10.1103/PhysRevD.77.046009 [arXiv:0712.4017 [hep-th]].
%608 citations counted in INSPIRE as of 26 Jan 2022



%\cite{Nojiri:2007cq}
\bibitem{Nojiri:2007cq}
S.~Nojiri and S.~D.~Odintsov,
%``Modified f(R) gravity unifying R**m inflation with Lambda CDM epoch,''
Phys. Rev. D \textbf{77} (2008), 026007
doi:10.1103/PhysRevD.77.026007 [arXiv:0710.1738 [hep-th]].
%352 citations counted in INSPIRE as of 26 Jan 2022














%\cite{Starobinsky:1980te}
\bibitem{Starobinsky:1980te}
A.~A.~Starobinsky,
%``A New Type of Isotropic Cosmological Models Without Singularity,''
Phys. Lett. B \textbf{91} (1980), 99-102
doi:10.1016/0370-2693(80)90670-X
%5351 citations counted in INSPIRE as of 26 Oct 2021



%\cite{Bezrukov:2007ep}
\bibitem{Bezrukov:2007ep}
F.~L.~Bezrukov and M.~Shaposhnikov,
%``The Standard Model Higgs boson as the inflaton,''
Phys. Lett. B \textbf{659} (2008), 703-706
doi:10.1016/j.physletb.2007.11.072 [arXiv:0710.3755 [hep-th]].
%1597 citations counted in INSPIRE as of 26 Oct 2021


%\cite{Hu:2007nk}
\bibitem{Hu:2007nk}
W.~Hu and I.~Sawicki,
%``Models of f(R) Cosmic Acceleration that Evade Solar-System Tests,''
Phys. Rev. D \textbf{76} (2007), 064004
doi:10.1103/PhysRevD.76.064004 [arXiv:0705.1158 [astro-ph]].
%1414 citations counted in INSPIRE as of 26 Oct 2021





%\cite{Bamba:2012qi}
\bibitem{Bamba:2012qi}
K.~Bamba, A.~Lopez-Revelles, R.~Myrzakulov, S.~D.~Odintsov and
L.~Sebastiani,
%``Cosmic history of viable exponential gravity: Equation of state oscillations and growth index from inflation to dark energy era,''
Class. Quant. Grav. \textbf{30} (2013), 015008
doi:10.1088/0264-9381/30/1/015008 [arXiv:1207.1009 [gr-qc]].
%61 citations counted in INSPIRE as of 26 Oct 2021

%\cite{Odintsov:2020vjb}
\bibitem{Odintsov:2020vjb}
S.~D.~Odintsov, V.~K.~Oikonomou, F.~P.~Fronimos and
K.~V.~Fasoulakos,
%``Unification of a Bounce with a Viable Dark Energy Era in Gauss-Bonnet Gravity,''
Phys. Rev. D \textbf{102} (2020) no.10, 104042
doi:10.1103/PhysRevD.102.104042 [arXiv:2010.13580 [gr-qc]].
%11 citations counted in INSPIRE as of 26 Oct 2021




%\cite{Odintsov:2020qyw}
\bibitem{Odintsov:2020qyw}
S.~D.~Odintsov, V.~K.~Oikonomou and F.~P.~Fronimos,
%``$f(R)$ gravity $k$-Essence late-time phenomenology,''
Phys. Dark Univ. \textbf{29} (2020), 100563
doi:10.1016/j.dark.2020.100563 [arXiv:2004.08884 [gr-qc]].
%8 citations counted in INSPIRE as of 26 Oct 2021



%\cite{Zhao:2008bn}
\bibitem{Zhao:2008bn}
G.~B.~Zhao, L.~Pogosian, A.~Silvestri and J.~Zylberberg,
%``Searching for modified growth patterns with tomographic surveys,''
Phys. Rev. D \textbf{79} (2009), 083513
doi:10.1103/PhysRevD.79.083513 [arXiv:0809.3791 [astro-ph]].
%231 citations counted in INSPIRE as of 30 Mar 2022


%\cite{Appleby:2009uf}
\bibitem{Appleby:2009uf}
S.~A.~Appleby, R.~A.~Battye and A.~A.~Starobinsky,
%``Curing singularities in cosmological evolution of F(R) gravity,''
JCAP {\bf 1006} (2010) 005 doi:10.1088/1475-7516/2010/06/005
[arXiv:0909.1737 [astro-ph.CO]].
%%CITATION = doi:10.1088/1475-7516/2010/06/005;%%
%163 citations counted in INSPIRE as of 27 Dec 2019




%\cite{Elizalde:2011ds}
\bibitem{Elizalde:2011ds}
E.~Elizalde, S.~D.~Odintsov, L.~Sebastiani and S.~Zerbini,
%``Oscillations of the F(R) dark energy in the accelerating universe,''
Eur. Phys. J. C \textbf{72} (2012), 1843
doi:10.1140/epjc/s10052-011-1843-7 [arXiv:1108.6184 [gr-qc]].
%37 citations counted in INSPIRE as of 28 Mar 2022



%\cite{Nadkarni-Ghosh:2021mws}
\bibitem{Nadkarni-Ghosh:2021mws}
S.~Nadkarni-Ghosh and S.~Chowdhury,
%``Non-linear density-velocity dynamics in $f(R)$ gravity from spherical collapse,''
doi:10.1093/mnras/stac133 [arXiv:2110.05121 [astro-ph.CO]].
%0 citations counted in INSPIRE as of 28 Mar 2022



%\cite{Nojiri:2003ni}
\bibitem{Nojiri:2003ni}
S.~Nojiri and S.~D.~Odintsov,
%``Modified gravity with ln R terms and cosmic acceleration,''
Gen. Rel. Grav. \textbf{36} (2004), 1765-1780
doi:10.1023/B:GERG.0000035950.40718.48 [arXiv:hep-th/0308176
[hep-th]].
%431 citations counted in INSPIRE as of 26 Jan 2022

%\cite{Bamba:2014mua}
\bibitem{Bamba:2014mua}
K.~Bamba, G.~Cognola, S.~D.~Odintsov and S.~Zerbini,
%``One-loop modified gravity in a de Sitter universe, quantum-corrected inflation, and its confrontation with the Planck result,''
Phys. Rev. D \textbf{90} (2014) no.2, 023525
doi:10.1103/PhysRevD.90.023525 [arXiv:1404.4311 [gr-qc]].
%42 citations counted in INSPIRE as of 26 Jan 2022







\end{thebibliography}
\end{document}